\documentclass{jpp}
\usepackage{graphicx}
\usepackage[utf8]{inputenc}
\usepackage[T1]{fontenc}
\usepackage{amsmath}
\usepackage{accents}
\usepackage[colorlinks=true]{hyperref}

\newtheorem{theorem}{Theorem}
\newtheorem{corollary}{Corollary}[theorem]

\shorttitle{Microstability of $\beta \sim 1$ tokamak equilibria}
\shortauthor{Gaur, Abel, Dickinson, and Dorland}

\title{Microstability of $\beta \sim 1$ tokamak equilibria}
\author{Rahul Gaur\aff{1}
  \corresp{\email{rgaur@umd.edu}},
  Ian G. Abel\aff{1},
 David Dickinson\aff{3},
 \and \newline William D. Dorland\aff{1, 2}}

\affiliation{\aff{1}Institute for Research in Electronics and Applied Physics, University of Maryland, College Park, MD 20740, USA
\aff{2} Department of Physics, University of Maryland, College Park, MD 20740, USA
\aff{3} Department of Physics, University of York, Heslington, York, YO10 5DD, UK}

\begin{document}

\maketitle

\begin{abstract}
High-power-density tokamaks offer a potential solution to design cost-effective fusion devices. One way to achieve high power density is to operate at a high $\beta$ value (the ratio of thermal to magnetic pressure), i.e., $\beta \sim 1$. However, a $\beta \sim 1$ state may be unstable to various pressure- and current-driven instabilities or have unfavorable microstability properties. To explore these possibilities, we generate $\beta \sim 1$ equilibria and investigate their stability.

Initially, we study an analytical technique that was used in the past to generate $\beta \sim 1$ equilibria and outline its limitations. Hence, we demonstrate the generation of high-$\beta$ equilibria with the computer code \texttt{VMEC}. We then analyze these equilibria to determine their stability against the infinite-$n$ ideal ballooning mode. We follow that by engaging in a detailed microstability study, beginning with assessments of electrostatic ITG and TEM instabilities. We observe interesting behavior for the high-$\beta$ equilibria -- stabilization of these modes through two distinct mechanisms. Finally, we perform electromagnetic gyrokinetic simulations and again observe stabilizing trends in the equilibria at high $\beta$.
 These trends are different from their lower $\beta$ counterparts and offer an alternative,  potentially favorable regime of tokamak operation.
\end{abstract}

\section{Introduction}
The most advanced fusion reactor designs are currently based on the tokamak concept.
These tokamak designs have historically led to large volume, large capital cost, plants such as ITER~\citep{progiterphysics1} and EU-DEMO~\citep{EUDEMO}. Improvements in these designs could be achieved by operating at high power density, with reduced plasma volume~\citep{menard2022fusion}.

The power density of a tokamak $P$ scales as $\beta^2$, where $\beta$ is the ratio of the plasma pressure to the magnetic pressure. Present-day tokamaks are low-$\beta$ devices. The achievable $\beta$ is typically limited by plasma instabilities. These can lead to disruptions or  large turbulent transport. The higher $\beta$ is, the higher the pressure and current are, and therefore the larger the free-energy available to drive these instabilities is. If these problems could be overcome, high-$\beta$ operation could be an attractive choice for future high power density~\citep{menard2022fusion} devices.

The high-beta, $\beta \sim 1$, regime has been explored previously in the context of asymptotic MHD equilibria by solving the Grad-Shafranov equation in the limit $\epsilon/(\beta q^2) \ll 1$~\citep{hsuartuncowley_highbeta}.
There have also been experimental explorations~\citep{gates2003high, sykes2000h} of high-$\beta$ operation of the MAST and START tokamaks. There have only been a few studies~\citep{hurricane2000internal, chance1990ideal} that investigated the process of accessing these states while maintaining ideal-MHD stability. Therefore, a detailed numerical analysis of these types of equilibria is required.

To that end, we generate a set of high-$\beta$ equilibria and study their susceptibility to local MHD and gyrokinetic instabilities. In context of local MHD we study stability of these equilibria to the infinite-$n$, ideal-ballooning mode. Then we perform linear gyrokinetic analyses of various electrostatic and electromagnetic modes of instability in these equilibria. These modes are known to cause significant heat and particle transport in existing devices~\citep{white2013multi, creely2017validation}

The remainder of this paper is divided as follows: in~\S\ref{sec:equilibrium}, we briefly describe the fundamentals of an axisymmetric equilibrium followed by the procedure devised by Hsu \textit{et al.} to solve the Grad-Shafranov equation. Next, we describe some of the limitations of their analytical equilibria. To ameliorate this, in~\S\ref{subsec:VMEC} we obtain numerical equilibria using the \texttt{VMEC} code~\citep{hirshman1983steepest_VMEC}. We describe and plot the equilibria for three different beta values: low-$\beta$ ($\beta \sim 0.01$), intermediate-$\beta$ ($\beta \sim 0.1$) and high-$\beta$ ($\beta \sim 1$) and two different triangularity values. In~\S\ref{sec:Ideal_ballooning_stability}, we introduce the technique devised by~\citet{greene1981second} to vary gradients of a local equilibria. We then use this to analyse the susceptibility of the chosen local equilibria to the ideal-ballooning instability and explain our observations. After ensuring ideal-ballooning stability for our high-$\beta$ equilibria, in~\S\ref{sec:microstability}, we briefly explain the linear, collisionless, gyrokinetic model and process for analyzing kinetic stability of local equilibria using the gyrokinetic code \texttt{GS2}.  In~\S\ref{sec:ITG-study}, we solve the gyrokinetic model for electrostatic fluctuations with adiabatic electrons and use the same techniques that we used in~\S\ref{sec:Ideal_ballooning_stability} to scan the growth rates of all the local equilibria with respect to the temperature gradient. In~\S\ref{sec:TEM-study}, we solve the gyrokinetic model for electrostatic fluctuations after relaxing the assumption of adiabatic electrons and use the same techniques as~\S\ref{sec:Ideal_ballooning_stability} to scan the growth rates of all the local equilibria with respect to the density gradient. In~\S\ref{sec:electromagnetic-GK}, we study the stability of local equilibria at their nominal gradients to electromagnetic fluctuations by solving the full gyrokinetic model and explain the advantages of negative-triangularity, high-$\beta$ equilibria. 
Finally in~\S\ref{sec:conclusions}, we summarise our work and discuss the directions in which it can be extended.
\section{Generating axisymmetric $\beta \sim 1$ equilibria}
\label{sec:equilibrium}
In this section, we will start with the general form of a divergence-free magnetic field, simplify it for an axisymmetric configuration and using the simplified form, obtain an explicit form of the steady-state momentum equation --- the Grad-Shafranov equation. In~\S\ref{sub:HAC_eqbm}, we discuss a general analytical technique that can be used to solve the Grad-Shafranov equation in the regime of large plasma pressure. In~\S\ref{sub:HAC_lim}, we explain the various limitations of solutions obtained from this approach and argue in favour of using a numerical solver (\texttt{VMEC})
instead. In the final section, we briefly explain how \texttt{VMEC} works and provide details for the equilibria generated for this work.

A divergence-free magnetic field $\boldsymbol{B}$ can be written in the Clebsch form~\citep{d2012flux}
\begin{equation}
    \boldsymbol{B} = \bnabla\alpha \times \bnabla \psi.
    \label{eqn:Div-free-B}
\end{equation}
We shall restrict our attention to solutions whose magnetic field lines lie on closed nested toroidal surfaces, known as flux surfaces. We label the flux surfaces by their enclosed poloidal flux $\psi$. On each flux surface, the line of constant $\alpha$, the field-line label, gives us the path of the magnetic field line. 

We will use the right-handed, cylindrical coordinate system $(R, \phi, Z)$ where $R$ and $Z$ are the radial and vertical distance from the origin and $\phi$ is the azimuthal angle about the symmetry axis. We also define a curvilinear coordinate system $(\psi, \phi, \theta)$ with $\psi$ being the flux surface label, $\phi$ being the cylindrical azimuthal angle and $\theta$ being the ``straight-field-line'' poloidal angle~\citep{d2012flux} such that $\alpha = \phi - q(\psi) \theta$ where
\begin{equation}
q(\psi) \equiv  \frac{d \chi}{d \psi} = \frac{1}{2\pi} \oint d\theta \, \frac{\boldsymbol{B}\bcdot \bnabla\phi}{\boldsymbol{B}\bcdot \bnabla \theta} = \frac{\boldsymbol{B}\bcdot\bnabla\phi}{\boldsymbol{B}\bcdot \bnabla \theta},
\label{eqn:Safety-factor-definition}
\end{equation}
is the safety factor, $\chi$ is the enclosed toroidal flux, and the line integral is over $\theta \in [0, 2\pi]$. The relation for $q(\psi)$ is consistent with the definition of $\alpha = \phi -  q\theta$ as can be seen by substituting $\alpha$ into $\eqref{eqn:Div-free-B}$ , obtaining
\begin{equation}
    \boldsymbol{B} = \bnabla\phi \times \bnabla\psi - q \bnabla \theta \times \bnabla\psi, 
    \label{eqn:Axisymmetric-B-0}
\end{equation}
and noting that $\eqref{eqn:Axisymmetric-B-0}$ satisfies $\eqref{eqn:Safety-factor-definition}$. For an axisymmetric $\boldsymbol{B}$, $\eqref{eqn:Axisymmetric-B-0}$ can be reduced further, to  
\begin{equation}
    \boldsymbol{B} = \bnabla\phi \times \bnabla\psi + F \bnabla\phi.
    \label{eqn:Axisymmetric-B}
\end{equation}
Here, $F = F(\psi, \theta)$. To generate an MHD equilibrium one has to solve the steady-state, ideal-MHD force balance equation~\citep{IdealMHD} which, in SI units, is
\begin{equation}
    \bnabla p = \frac{(\bnabla\times \boldsymbol{B})\times \boldsymbol{B}}{\mu_0} = \frac{\boldsymbol{B}\bcdot \bnabla \boldsymbol{B}}{\mu_0}-\bnabla\left(\frac{B^2}{2 \mu_0}\right).
    \label{eqn:ideal-MHD-force-balance}
\end{equation}
For an axisymmetric system, we can substitute the form of $\boldsymbol{B}$ from~\eqref{eqn:Axisymmetric-B} and choose the toroidal component of~\eqref{eqn:ideal-MHD-force-balance} to get $\boldsymbol{B}\bcdot \bnabla F = 0$ which implies $F = F(\psi)$. Upon choosing the radial component, we obtain the Grad-Shafranov~\citep{grad1958proceedings, shafranov1957equilibrium} equation
\begin{equation}
\boldsymbol{\Delta}^{*}\psi \equiv R^2 \bnabla\bcdot \left(\frac{\bnabla \psi}{R^2}\right) = -\mu_0 R^2 \frac{dp}{d\psi} - F\frac{dF}{d\psi}.
\label{eqn:Grad-Shafranov-equation}
\end{equation}
The Grad-Shafranov equation is a non-linear equation for the poloidal flux $\psi(R, Z)$ that depends on the pressure $p(\psi)$ and current $F(\psi)$ profiles. As discussed in the introduction, the fusion power output $P$ of a tokamak scales as $\beta^2$ --- $\beta \equiv p/(B^2/(2\mu_0))$ is the ratio of plasma pressure to magnetic pressure --- which makes high-$\beta$ operation an attractive concept. Therefore, our objective is to analyze the stability of equilibria that have $\beta \sim 1$. In~\S\ref{sub:HAC_eqbm}, we will examine the analytical work done by Hsu, Artun and Cowley~\citep{hsuartuncowley_highbeta} to generate such equilibria.  

\subsection{Hsu, Artun and Cowley's ($\beta \sim 1$) equilibria} 
\label{sub:HAC_eqbm}
The Grad-Shafranov equation has been solved analytically~\citep{shafranov1957equilibrium,atanasiu2004analytical} for different pressure and current profiles. By design, analytical solutions are faster to compute than their numerical counterparts and do not suffer from resolution or convergence related problems. Therefore, multiple approaches have been taken to solve equation $\eqref{eqn:Grad-Shafranov-equation}$ analytically in the limit $\beta \sim 1$. 

\citet{cowley1991analytic_highbeta} solved the Grad-Shafranov equation for $\beta \sim 1$ equilibria for large-aspect ratio tokamaks and later analyzed their ideal ballooning and interchange stability~\citep{cowley1991stability_highbeta}. The most general analytical theory to generate $\beta \sim 1$ equilibria was first developed by $\textrm{Hsu, Artun and Cowley}$~\citep{hsuartuncowley_highbeta} where the equation is solved analytically in the limit
\begin{equation}
\delta_{\mathrm{Hsu}} \equiv \sqrt{\epsilon/(\beta q^2)} \ll 1,
\label{eqn:delta-Hsu}
\end{equation}
where $\epsilon = a/R_0$ is the aspect ratio of the flux surface --- $a$ being the minor radius of a flux surface and $R_0$ being the radial distance of the magnetic axis from the symmetry axis. For these equilibria, it is assumed $F/(R B_{\mathrm{p}}) \sim q/\epsilon$ where  $B_{\mathrm{p}} = \boldsymbol{B}\bcdot\bnabla \theta/|\bnabla \theta|$ is the poloidal field, and we assume $\epsilon \sim 1$ for all the surfaces of interest. Given these assumptions, the Grad-Shafranov equation on the inboard side becomes
\begin{equation}
    -\underbrace{\frac{d}{d\psi}(\mu_0 p)}_{\textit{O}(\beta)}  -\underbrace{\frac{d}{d\psi}\left(\frac{F^2}{2R^2}\right)}_{\textit{O}(1)} + \underbrace{\frac{F^2}{R^3}\frac{dR}{d\psi}}_{\textit{O}(\epsilon \beta)} = 0.
    \label{eqn:Core-equation}
\end{equation}
The symbol $\textit{O}$ beneath each term denotes its size relative to the middle term. Physically, the equation implies that the difference in the plasma and magnetic pressure is balanced by toroidal curvature. Note that the term on the left-hand side in $\eqref{eqn:Grad-Shafranov-equation}$
\begin{equation}
    \bnabla \bcdot \left(\frac{\bnabla \psi}{R^2}\right) \sim \textit{O}\left(\frac{\epsilon^2}{q^2}\right),
\end{equation}
is dropped as it is $\textit{O}(\delta_{\mathrm{Hsu}})$ smaller than the rest of the terms.  Following Hsu \textit{et al.}, we introduce a function $\hat{R} = \hat{R}(\psi) = R(\psi, Z = 0)$ such that
\begin{equation}
\mu_0 \hat{R}(\psi)^2 \frac{dp}{d\psi} = F\frac{dF}{d\psi},
\label{eqn:lowest-order-Hsu}
\end{equation}
using which we can write $\eqref{eqn:Core-equation}$ as
\begin{equation}
    [\hat{R}(\psi)-R^2]\frac{dF^2}{d\psi} = 0.
    \label{eqn:Core-solution}
\end{equation}
Equation $\eqref{eqn:Core-solution}$ implies that the lowest-order solution is a vertical line in the $(R, Z)$ plane for each value of $\psi$. Hsu \textit{et al}. refer to this as the ``core'' solution. Clearly, $R = \hat{R}$ does not give us closed flux surfaces. This indicates the existence of boundary layer solution that includes contributions from the operator $\boldsymbol{\Delta}^{*}\psi$.

\begin{figure}
\centering
\includegraphics[width=0.92\textwidth, trim={0 {0.011\textwidth} 0 0}, clip]{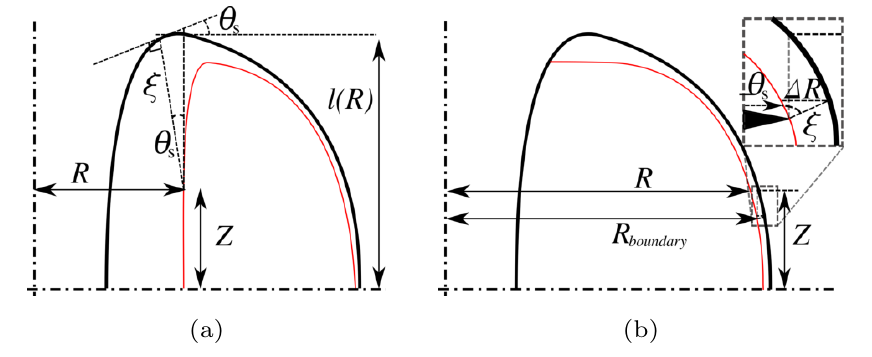}
\caption{This figure illustrates the process of creating the two branches of the $\beta \sim 1$ solution in Hsu \textit{et al.} This construction is used to obtain equation $\eqref{eqn:Core_and_BL_solution}$. The bold black line is the LCFS and red line is the flux surface contour --- (\textit{a}) shows the core solution which is only a good approximation on the inboard side of the device and (\textit{b}) shows the boundary layer solution which is only valid in the boundary layer region. The inset in the right figure highlights the approximation $\xi(\hat{R}, R) = \xi(\hat{R}, R_{\mathrm{boundary}})$ which is necessary to obtain the second equation in $\eqref{eqn:Core_and_BL_solution}$.}
\label{fig:Hsu-et-al-illustration}
\end{figure}

To solve the equation in the boundary layer region, Hsu \textit{et al.} move to the coordinate system $(\xi, \eta, \phi)$ where $\xi$ is the perpendicular distance to a point on a flux surface from the Last Closed Flux Surface (henceforth LCFS), $\phi$ is the toroidal angle and $\eta$ is the third curvilinear coordinate. In this coordinate system, $\eqref{eqn:Grad-Shafranov-equation}$ becomes
\begin{equation}
    \mu_0 (\hat{R}^2 - R^2) \frac{dp}{d\psi} = \frac{\partial^2 \psi}{\partial \xi^2} + \textit{O}\left(\frac{ \delta_{\mathrm{Hsu}}}{\epsilon}\right),
    \label{eqn:BL-equation}
\end{equation}
where the gradient $\partial/\partial \xi \sim \textit{O}(1/\delta_{\mathrm{Hsu}})$ in the boundary layer and the gradient $\partial/\partial \eta  \sim \textit{O}(1) \ll \partial/\partial \xi$. This makes left-hand-side term $\textit{O}(\beta q^2/\epsilon^2)$ and the right-hand-side term $\textit{O}(1/(\epsilon \delta_{\mathrm{Hsu}}^2))$ in terms of the poloidal field $B_{\mathrm{p}}$. Using the definition of $\delta_{\mathrm{Hsu}}$ in $\eqref{eqn:delta-Hsu}$, we can see that both the terms in~$\eqref{eqn:BL-equation}$ have the same order. Equation~$\eqref{eqn:BL-equation}$ was solved by Hsu \textit{et al.} to obtain
\begin{equation}
    \xi = \int_{R_{\mathrm{min}}}^{\hat{R}^{'}}\frac{(d\psi/d\hat{R}^{'}) d\hat{R}^{'}}{\sqrt{\int_{R}^{\hat{R}^{'}} \mu_0({\hat{R}^{''}}^2 - R^2) \frac{dp}{d\hat{R}^{''}}   d \hat{R}^{''} }}.
    \label{eqn:xi-final-formula}
\end{equation}
After obtaining the value of $\xi(\hat{R}, R)$ one creates solutions valid in the two regions, the core solution, expressed as a function $Z = Z(\hat{R}, R)$ and the boundary layer solution $R = R(\hat{R}, Z)$ such that
\begin{equation}
\begin{gathered}
    Z(\hat{R}, R) = l(R) - \frac{\xi(\hat{R}, R)}{\cos(\theta_{\mathrm{s}})}, \quad  \theta_{\mathrm{s}} = \arctan\left( \frac{dl}{dZ}\right)\\
    \Delta R = R_{\mathrm{boundary}}(Z) - R(\hat{R}, Z)  =  \frac{\xi(\hat{R}, R_{\mathrm{boundary}})}{\sin(\theta_{\mathrm{s}})},
    \label{eqn:Core_and_BL_solution}
\end{gathered}
\end{equation}
 are joined together to form a closed flux surface $\psi = \psi(R, Z)$. In $\eqref{eqn:Core_and_BL_solution}$, $l(R)$ is the vertical distance of a point on the boundary from the horizontal axis and $R_{\mathrm{boundary}}$ is the radial distance of a point on the boundary from the symmetry axis. The angle $\theta_{\mathrm{s}}$ is the angle subtended by the tangent at a point on the boundary with the horizontal axis, the angle of the slope. The illustration shown in figure~\ref{fig:Hsu-et-al-illustration} explains the process of obtaining $\eqref{eqn:Core_and_BL_solution}$.

\begin{figure}
\centering
\includegraphics[width=\textwidth, trim={0 {0.011\textwidth} 0 0}, clip]{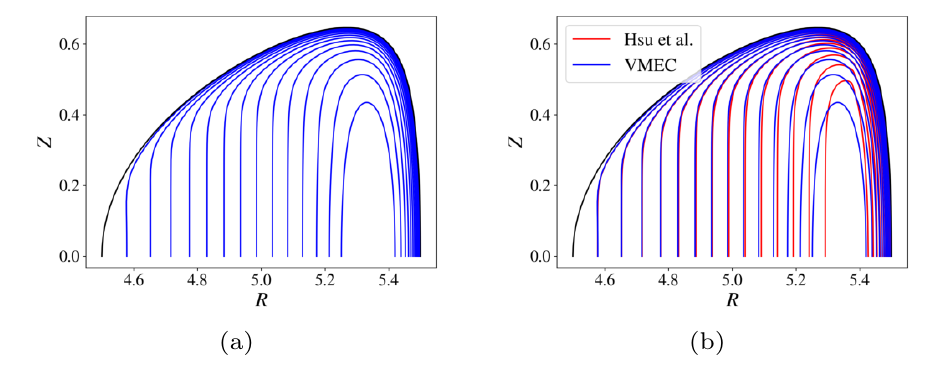}
\caption{This figure shows (\textit{a}) the numerical equilibrium solution and (\textit{b}) a comparison between the analytical equilibrium from figure $6$ in the paper by Hsu \textit{et al.} and the same equilibrium generated using \texttt{VMEC}. We can clearly see the significant deviation of the analytical solution and how it develops a kink near the outboard side as we approach the magnetic axis.}
\label{fig:Hsu-VMEC-comparison}
\end{figure}

To create a $\beta \sim 1$ solution using this procedure, one needs user-defined, analytically integrable input functions $1/\hat{R}(d\psi/d\hat{R})$ and $p = p(\hat{R})$ for the numerator and denominator in \eqref{eqn:xi-final-formula}, respectively. After obtaining $\xi$ from \eqref{eqn:xi-final-formula}, one uses the boundary shape $Z = l(R)$ and $\eqref{eqn:Core_and_BL_solution}$ to get the respective solution segments. In their work, Hsu \textit{et al.} choose rational polynomial functions for all three input profiles. Exact forms of these functions are given in appendix~\ref{app:Hsu_et_al_profiles}. To understand the differences between the analytical and numerical solutions, we reproduce a high-$\beta$ equilibrium from figure $6$ in Hsu \textit{et al.} and compare it with the corresponding \texttt{VMEC} equilibrium in figure~\ref{fig:Hsu-VMEC-comparison} of this paper.
While the procedure of Hsu \textit{et al.} works well for generating the contours of flux surfaces $\psi$ close to the edge, we will show that the quantities needed for doing a local stability analysis can be discontinuous and deviate significantly from the exact numerical solution. We elucidate these issues in the next section.

\subsection{Limitations of the Hsu \textit{et al.} equilibria}
\label{sub:HAC_lim}
In writing~$\eqref{eqn:Core-equation}$, we assumed $\epsilon \sim \textit{O}(1)$ and the poloidal pressure gradient to be negligible in the core. This assumption does not hold for surfaces close to the magnetic axis. Indeed, the surfaces near the magnetic axis are ellipses~\citep{IdealMHD} which will deviate significantly from a straight line solution predicted by $\eqref{eqn:Core-solution}$. The distinction between the ``core'' and the boundary layer becomes less apparent as one moves towards the magnetic axis. Finally, one cannot ensure that the $\xi$ for surfaces close to the magnetic axis is a unique coordinate; there could be multiple points on the boundary corresponding to the same point on the flux surface. The uniqueness can be strictly guaranteed globally only for a perfectly circular boundary.

The boundary layer solution suffers from multiple issues as well. First and foremost, Hsu \textit{et al.} use an approximation $\xi(\hat{R}, R) = \xi(\hat{R}, R_{\mathrm{boundary}})$ for the boundary layer solution which only holds for surfaces that are close to the LCFS. As one moves towards the magnetic axis, the surfaces will move away from the LCFS and affect the accuracy of the solution. Moreover, in $\eqref{eqn:BL-equation}$, once the boundary layer width becomes large, the expression in $\eqref{eqn:Core_and_BL_solution}$ does not take into account the higher-order corrections to $\xi$ and fails to provide the requisite accuracy.

\begin{figure}
    \centering
    \includegraphics[width=0.4\linewidth, trim={0 0 0 0}, clip]{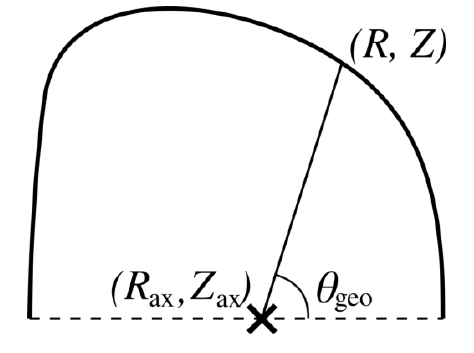}
    \caption{This figure illustrates the definition of $\theta_{\mathrm{geo}}$. The coordinates of the magnetic axis, marked with a cross, are $(R_{\mathrm{ax}}, Z_{\mathrm{ax}})$.}
    \label{fig:theta_geo}
\end{figure}

\begin{figure}
\centering
\includegraphics[width=0.97\textwidth, trim={0 {0.011\textwidth} 0 0}, clip]{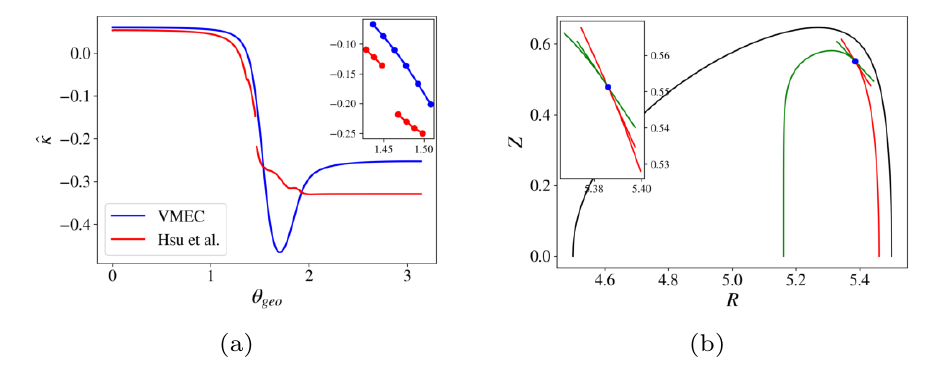}
\caption{In this figure (\textit{a}) compares $\hat{\kappa}$ vs. $\theta_{\mathrm{geo}}$ obtained using the analytical equilibrium(in figure $6$) in Hsu \textit{et al.} with the corresponding \texttt{VMEC} equilibrium for a common flux surface(shown in (\textit{b})). The inset plot in (\textit{a}) shows a zoomed-in version of the same plot near the discontinuity at $\theta_{\mathrm{geo}} = 1.46$. Figure (\textit{b}) shows the difference in slopes of the tangents at the point of discontinuity, with a zoomed-in version in the inset. Notice also the deviation of $\hat{\kappa}$ in (\textit{a}) for $\theta_{\mathrm{geo}}>1.5$. There are other issues like the small sharp feature near $\theta_{\mathrm{geo}} = 1.3$ that we will not delve into.}
\label{fig:cvdrift_compare}
\end{figure}

Combining the two solutions causes discontinuity in physical quantities due to the presence of sharp gradients. To understand this problem, we plot a physical quantity that arises in the both the ballooning and gyrokinetic equation that can be seen in $\eqref{eqn:ballooning-equation}$ and \eqref{eqn:particle-drift} in~\S\ref{sec:Ideal_ballooning_stability} and~\S\ref{sec:microstability}, respectively--- as a curvature drive in the former and as a component of the curvature drift in the latter. The quantity is
\begin{equation}
    \hat{\kappa} = \frac{1}{B^3} \left(\boldsymbol{b}\times\boldsymbol{\kappa}\right) \bcdot \bnabla\alpha, \quad \boldsymbol{\kappa} = (\boldsymbol{b}\bcdot\bnabla\boldsymbol{b}),
\end{equation}
where $\boldsymbol{\kappa}$ is the field-line curvature. To better understand the discontinuity problem, we plot $\hat{\kappa}$ in figure~\ref{fig:cvdrift_compare} as a function of the geometric poloidal angle defined in the figure~\ref{fig:theta_geo}.

 The coordinate $\theta_{\mathrm{geo}}$ is advantageous as it, unlike $\theta$, is a physically intuitive poloidal angle. Mathematically, $\theta_{\mathrm{geo}}$ is a monotonic function of $\theta$ --- $\hat{\kappa}(\theta_{\mathrm{geo}})$ can always be transformed to $\hat{\kappa}(\theta)$ and vice-versa.  To emphasize the importance of smoothness, we also plot the tangents on the flux surface on either side of the point at which $\hat{\kappa}$ is discontinuous. 
The kink in $\psi$ seen in figure~$4$(\textit{b}) manifests itself as a discontinuity in quantities like $\boldsymbol{b}$, $\bnabla\psi$ and $\bnabla\alpha$. This causes the geometric factors, and hence the physical quantities needed for a local stability analysis to become discontinuous. Furthermore, in the regions where the gradients are continuous, for reasons mentioned at the beginning of this subsection, the distances between the surfaces deviate from the exact equilibrium, especially as $\theta_{\mathrm{geo}}>1.5$. To alleviate these problems, we will use the \texttt{VMEC} code to generate the equilibria used in this study. 

\subsection{Numerical equilibria}
\label{subsec:VMEC}
We generate equilibria numerically using the 3-D equilibrium code \texttt{VMEC}~\citep{hirshman1983steepest_VMEC}. \texttt{VMEC} works by minimizing the integral 
\begin{equation}
W = \int\left(\frac{p}{\gamma -1} + \frac{B^2}{2\,\mu_0}\right) dV,
\label{eqn:energy_integral}
\end{equation}
subject to multiple constraints, which for axisymmetric equilibria is equivalent to solving the Grad-Shafranov equation~\citep{kruskal1958equilibrium}. For our study, \texttt{VMEC} takes the shape of the boundary surface along with the global radial pressure $p(s)$ and safety factor $q(s)$ or enclosed toroidal current $G(s)$ as a function of the normalized toroidal flux $s$. It then creates flux surfaces to minimize the integral in~$\eqref{eqn:energy_integral}$ on each surface for a fixed $p$ and $q$ (or $G$). We choose the safety factor instead of toroidal current as it varies slowly~\citep{flowtome2-electrons} compared to the plasma currents in the limit of small electron-to-ion mass ratio.

A comparison between the equilibrium given in figure $6$ of Hsu \textit{et al.} and the corresponding numerical equilibrium is given in figure~\ref{fig:Hsu-VMEC-comparison}. We can see that the \texttt{VMEC} solution is smooth, unlike the Hsu \textit{et al.} solution. Hence, it should be possible to do stability analysis of the chosen equilibria reliably using \texttt{VMEC}. 

In the following paragraphs, we will explain the process of generating the data for this study using \texttt{VMEC}.  For each \texttt{VMEC} equilibrium, we pick two different radially local regions --- equilibria that satisfy $\eqref{eqn:Grad-Shafranov-equation}$ and are localized to a flux surface. It is important to point out that for most of this paper, we will only investigate modes that have small wavelengths perpendicular to the field line, i.e., modes that are localized to a flux surface, henceforth referring to our study as a local stability analysis. Since our aim is to have ample variability in our input data, at the end of this section we justify our choices by looking at the chosen flux surfaces and their corresponding $\beta$ values.

\begin{figure}
\centering
\includegraphics[width=0.9\textwidth, trim={0 {0.011\textwidth} 0 0}, clip]{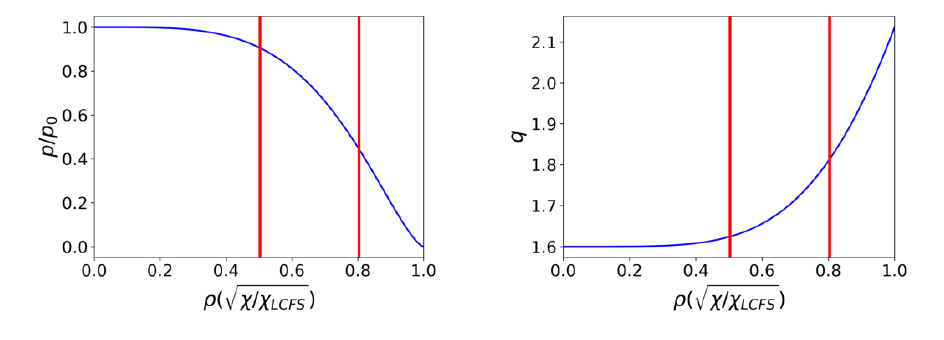}
\caption{This figure shows the safety factor and normalized pressure profiles used for creating the equilibria. The two red lines correspond to the values of the normalized radius $\rho$ at which the local equilibria will be analyzed for their stability}
\label{fig:pressure-and-safety-factor-profile}
\end{figure}

We produce high-radial-resolution equilibria using the fixed-boundary solver in \texttt{VMEC} after providing it with an L-mode like pressure profile $p = p(s)$ and a monotonic safety factor $q = q(s)$ profile as a function of the normalized toroidal flux $s = \chi/\chi_{\mathrm{LCFS}}$ and the LCFS shape. We choose a simple form for the profiles $p = p(s)$ and $q = q(s)$ given by
\begin{equation}
\begin{gathered}
    p = n T, \quad p_0 = n_0 T_0 \\
    n(s) = n_0(1 + \nu_n)(1-s^2)^{\nu_n},\quad T(s) = T_0(1 + \nu_T)(1-s^2)^{\nu_T}\\
    q = q_0(1+s^{2 \nu_q})^{1/(2 \nu_q)}.
\end{gathered}
\label{eqn:input_params}
\end{equation}
The different parameters are given in the table~\ref{tab:Table-1}
\begin{table}
  \begin{center}
\def~{\hphantom{0}}
  \begin{tabular}{lcccccc}
    $n_0(m^{-3})$ &  $\nu_n$  & $T_0(eV)$ & $\nu_T$ & $q_0$ & $\nu_q$ & $\chi_{\mathrm{LCFS}}(T-m^2)$\\[3pt]
    $5 \times 10^{20}$  & $0.4$  & $10\, \tilde{p}_0$ & $1.1$ & $1.6$ & $1.2$ & $1.0$ \\
  \end{tabular}
    \caption{VMEC equilibria input parameters}
  \label{tab:Table-1}
  \end{center}
\end{table}
\noindent
where $\tilde{p}_0 \in [0, 10, 70]$ for the low, intermediate and high-beta equilibria, respectively. We also choose two different LCFS shapes described by a Miller parameterization~\citep{miller1998noncircular}
\begin{equation}
\begin{gathered}
    R = R_0 + a \cos(t + (\sin^{-1}\delta)\sin t)\\
    Z = a \kappa \sin(t)
    \label{eqn:Miller_parametrization}
\end{gathered}
\end{equation}
for each. The parameter $t$ varies from $[0, 2 \pi)$. The values of the rest of the parameters in $\eqref{eqn:Miller_parametrization}$ are given in  table~\ref{tab:Table-2}.
\begin{table}
\begin{center}
\begin{tabular}{lccc}
 $R_0(m)$ &  $a(m)$  & $\delta$ & $\kappa$\\[3pt]
 $1.6$  & $0.6$  & $\pm 0.4$ & $1.3$\\
\end{tabular}
\end{center}
\caption{Miller Parameters for the outer boundary}
\label{tab:Table-2}
\end{table}
\noindent
The radial coordinate that we will use for all the stability analyses is $\rho = \sqrt{s} = \sqrt{\chi/\chi_{\mathrm{LCFS}}}$ since it is a better measure of the radial distance from the magnetic axis than the normalized poloidal flux $\psi/\psi_{\mathrm{LCFS}}$. The safety factor and pressure profiles  as a function of $\rho$ are given in figure~\ref{fig:pressure-and-safety-factor-profile}.
For all our studies, we use the same safety factor $q$ and the normalized pressure profile $p$ with different values of $\tilde{p}_0$. In this way, we are able to create three different pressure profiles with on-axis $\beta \sim 0.01, 0.1, 1$ corresponding to $\tilde{p}_0 = 1, 10, 70$, respectively. Henceforth, we will refer to the equilibria with $\tilde{p}_0 = 1, 10, 70$ as low, intermediate and high-$\beta$ or $\beta \sim 0.01, \beta \sim 0.1$ and $\beta \sim 1$, respectively.
We need to pick flux surfaces for our local stability analyses. In this study, we choose surfaces at normalized radii $\rho =  0.5$ and $0.8$. In total, there are twelve local equilibria in our study: $3$ $\beta$ values $\times$ $2$ boundary shapes $\times$ $2$ $\rho$ values. Because $\beta$ varies over a flux surface, it will be convenient to introduce a reference magnetic field for each global equilibrium and redefine
\begin{equation}
    \beta(\rho) = 2 \mu_0 p(\rho)/B_{\mathrm{N}}^2,
\end{equation}
\begin{equation}
     B_{\mathrm{N}} = \chi_{\mathrm{LCFS}}/(\pi a_{\mathrm{N}}^2),
\end{equation}
where $B_{\mathrm{N}}$ is a reference magnetic field and $a_{\mathrm{N}}$ is the effective minor radius such that $\pi a_{\mathrm{N}}^2$ is equal to the area enclosed by the boundary and $\chi_{\mathrm{LCFS}}$ is the toroidal flux enclosed by the LCFS. For this study, $a_{\mathrm{N}} = 0.684 \,m, B_{\mathrm{N}} = 0.681\, T$. The  values of $\beta$ obtained from \texttt{VMEC} are given in table~\ref{tab:Table-3}. 
\begin{table}
\centering
\begin{tabular}{ccccc}
 $\delta$ & $\rho$ & Low-$\beta$ & Intermediate-$\beta$ & High-$\beta$\\[3pt]
 $0.4$ &  $0.5$  & $0.011$  & $0.11$ & $0.77$ \\
 $0.4$ &  $0.8$  & $0.006$  & $0.064$ & $0.45$ \\
 $-0.4$ & $0.5$ & $0.011$  & $0.11$ & $0.77$ \\
 $-0.4$ &  $0.8$ & $0.006$  & $0.064$ & $0.45$ \\
\end{tabular}
\caption{Reference $\beta$ values for selected surfaces}
\label{tab:Table-3}
\end{table}

\begin{figure}
    \centering
    \includegraphics[width=1.01\textwidth, trim={0 {0.011\textwidth} 0 0}, clip]{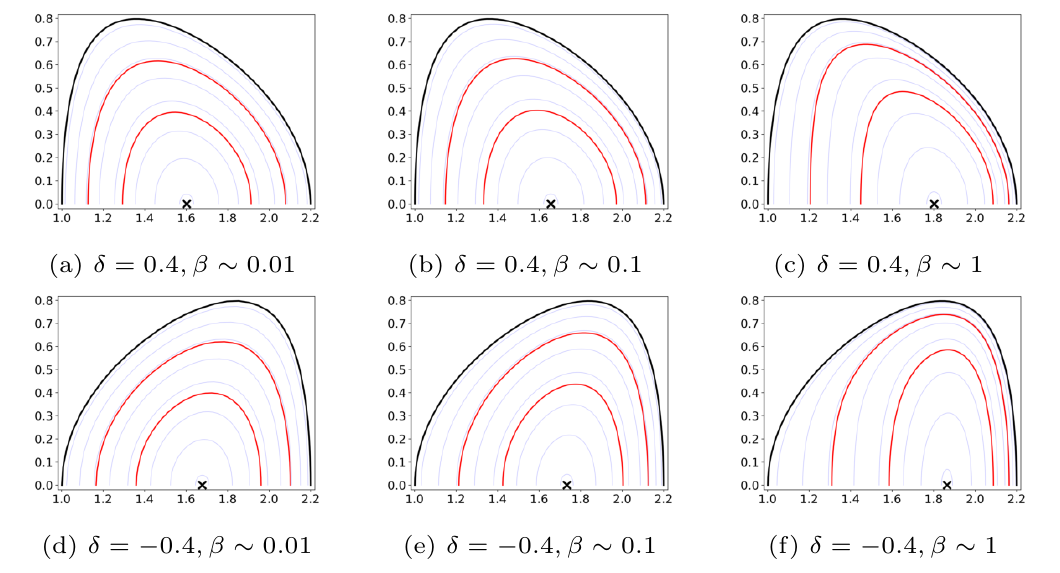}
\caption{This figure shows the flux surfaces for all the equilibria generated using \texttt{VMEC}. The local equilibria that will be studied in this paper are highlighted in red. The magnetic axis in each figure is the black cross.}
\label{fig:VMEC-equilibria}
\end{figure}

Each equilibrium has $512$ surfaces with each surface represented by $40$ poloidal modes. We found that the equilibria were converged with this choice of resolution. All the equilibria that we investigate in this study are up-down symmetric which is why we only show the upper half. The flux surface contours for the twelve equilibria are shown in figure~\ref{fig:VMEC-equilibria}.

\begin{figure}
    \centering
    \includegraphics[width=0.95\textwidth]{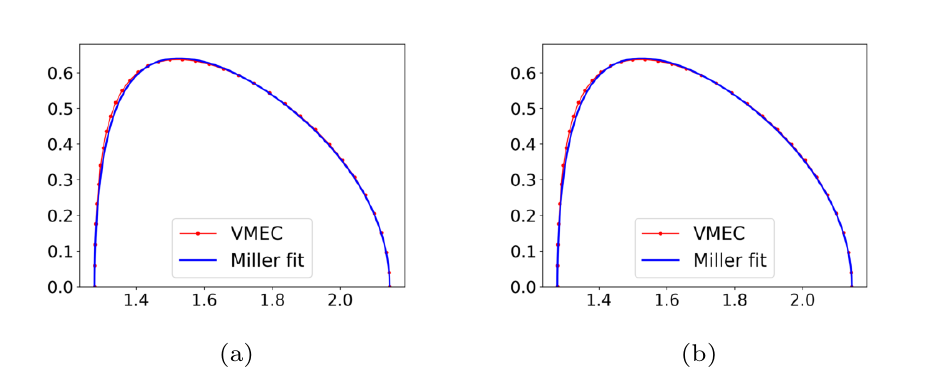}\\[-6pt]
\caption{This figure shows two high-$\beta$ equilibria and their corresponding best-Miller-fit. We can see that the fit for the negative triangularity is worse due to the ``squareness'' of the flux surface on the inboard side. The agreement between gradients of various physical quantities will be even worse.}
\label{fig:VMEC-Miller-fit}
\end{figure}

The numerical high-$\beta$ equilibria show qualitative features like the vertical ``core'' and thin boundary layer as shown by Hsu \textit{et al.} It is interesting to see that the negative triangularity equilibria high-$\beta$ equilibria are more strongly shaped than the positive triangularity ones --- the vertical core solution causes the inboard side to develop a ``squareness''. We illustrate the ``squareness'' in figure~\ref{fig:VMEC-Miller-fit}.
Most importantly, these numerical equilibria do not suffer from any of the issues discussed in the previous subsection (as seen in figure~\ref{fig:cvdrift_compare}). Therefore, all the resulting geometric coefficients are smooth which allows the local stability analyses in the following sections.

\section{Infinite-$n$ ideal-ballooning stability}
\label{sec:Ideal_ballooning_stability}
In this section, we will investigate the equilibria generated in the previous sections for their  stability to ideal ballooning modes. In~\S\ref{subsec:Phy-and-math-description}, we describe the physics basis and mathematical formulation of the ideal-ballooning problem. In~\S\ref{subsec:Greene-Chance}, we review the tools for locally varying  equilibria and introduce the concept of an $\hat{s}-\alpha_{\mathrm{MHD}}$ analysis. In the final section, we present the results from our study and discuss their implications.    
\subsection{Physical and mathematical description}
\label{subsec:Phy-and-math-description}
One of the most important MHD instability for us to investigate is the ideal-ballooning instability \citep{ConnorHastieTaylorProcRoySoc} --- a field-aligned, pressure-driven Alfv\'{e}n wave that grows when the destabilizing pressure gradient in the region of ``bad'' curvature exceeds the stabilizing effect of field-line bending. The region of ``bad'' curvature is a region of a flux surface where $\boldmath{\kappa} \cdot \bnabla p< 0$, such that the field line curvature is in a direction opposite to the plasma pressure. For most tokamak equilibria, this region lies on the outboard side.

The equation governing the ideal-ballooning mode can be obtained by minimizing the ideal-MHD energy integral~\citep{bernsteinenergyprinciple} for incompressible modes in the limit of large toroidal mode number. Doing so gives us a differential equation that determines $X$, the radial displacement of said mode along a field line. To ensure that the displacement $X$ satisfies the periodicity condition on the surface of interest and nearby surfaces, one uses the ballooning transformation
\begin{equation}
    X =  \sum_{N = -\infty}^{\infty} \hat{X}(\theta - 2\pi N)\, e^{in(\phi - q (\theta- \theta_0 - 2\pi N)))}, \quad N \in \mathbb{Z},
\end{equation}
subject to the condition
\begin{equation}
    \lim_{\theta \rightarrow \pm \infty} \hat{X}(\theta; \psi, \theta_0) = 0,
\end{equation}
and solves for $\hat{X}$. The variable $n$ is the toroidal mode number, $\theta_0$ is the ballooning parameter\footnote{In the context of infinite-$n$ ideal ballooning mode analyses, there is a value of the ballooning parameter $\theta_0$ at which the ballooning mode is the least stable. To find this value, one treats $\theta_0$ as a parameter and scans over its values to find the $\theta_0$ for which $\omega^2$ is the smallest.} and the rest of the terms are defined in~\S\ref{sec:equilibrium}. Upon minimizing the ideal-MHD energy integral and using the ballooning transformation, one obtains the ideal ballooning equation~\citep{ConnorHastieTaylorProcRoySoc, DewarGlasserballooning}   
\begin{equation}
    \frac{1}{\mathcal{J}}\frac{\partial}{\partial \theta}\left( \frac{\lvert \bnabla\alpha \rvert^2 }{\mathcal{J}\, B^2} \frac{\partial \hat{X}}{\partial \theta}\right) + 2 \frac{dp}{d\psi}\left[ \boldsymbol{B}\times\bnabla\left(p + \frac{B^2}{2}\right)\bcdot\bnabla{\alpha}\right] \hat{X} = \rho \omega^2 \frac{\lvert\bnabla\alpha\rvert^2}{B^2} \hat{X}, 
    \label{eqn:ballooning-equation}
\end{equation}
where $\rho$ is the plasma mass density and $\hat{X} = \hat{X}(\theta; \psi, \theta_0)$ is the eigenfunction in ballooning space and $\omega^2$ is the eigenvalue.
The ballooning equation balances the stabilizing field-line bending term and destabilizing pressure gradient with the inertia of the resulting Alfv\'{e}n wave, oscillating with a frequency $\omega$. Note that~\eqref{eqn:ballooning-equation} depends on $\psi$ only as a parameter and we can compute the coefficients from the on-surface equilibrium quantities and their first derivatives. Therefore, it is possible to study the ballooning stability of the local equilibria that we chose in the previous section.

Due to the self-adjoint nature of ideal-MHD, all the eigenvalues ($\omega^2$) of equation $\eqref{eqn:ballooning-equation}$ will be real numbers. Hence, $\omega$ will either be purely real, an oscillating mode or purely imaginary, a growing mode. We refer to the oscillating modes as stable and growing modes as unstable. Since unstable modes are of more significance to us, we will plot the normalized growth rate when plotting the results in~\S\ref{subsec:ideal-ballooning-results}. We define the normalized growth rate as follows 
\begin{equation}
 \gamma = -i\, \omega a_{\rm{N}}/v_{\mathrm{th, i}},   
 \label{eqn:ideal-ballooning-growth-rate-defn}
\end{equation}
where $v_{th,i} = \sqrt{2T_{\mathrm{i}}/m_{\mathrm{i}}}$ is the ion-thermal velocity and $T_{\mathrm{i}}$ and $m_{\mathrm{i}}$ are the temperature and mass of the ions, respectively.

\subsection{The Greene-Chance analysis}
\label{subsec:Greene-Chance}
\begin{figure}
\centering
{\includegraphics[width=0.5\textwidth]{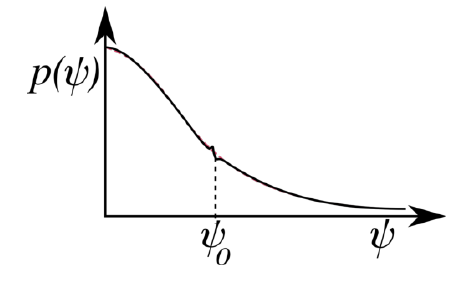}}\\[-3pt]
\caption{This figure summarizes the idea of Greene and Chance. The new pressure profile (black) with localized variation over the flux surface $\psi = \psi_0$ lies over the equilibrium profile (dashed red). Even though the variation in pressure at $\psi = \psi_0$ is small, the change in the pressure \textit{gradient} can be large.}
\label{fig:Greene-Chance-illustration}
\end{figure}

To better understand the stability of a local equilibrium, we need the ability to vary that equilibrium. This can be done by changing the magnetic shear and pressure gradient independently about their nominal values --- equivalent to varying the current and pressure gradient --- quantities that determine the solution to ~\eqref{eqn:Grad-Shafranov-equation}. This gives us the ability to generate multiple local equilibria satisfying the Grad-Shafranov equation and do a stability analysis without recalculating the global equilibrium. We define
\begin{align}
    \hat{s} &= \frac{\rho}{q}\frac{dq}{d\rho},\\
    \alpha_{\mathrm{MHD}} &= -\frac{2 \rho\, q^2}{\epsilon\, B_{\mathrm{N}}^2}\frac{dp}{d\rho} \label{eqn:alpha_MHD}
\end{align}
 as the magnetic shear and pressure gradient, respectively. This method of varying a local equilibrium through $\hat{s}$ and $\alpha_{\mathrm{MHD}}$ is known as an $\hat{s}-\alpha_{\mathrm{MHD}}$ analysis. This technique was developed by~\citet{greene1981second} and has been used extensively to study local stability of different axisymmetric equilibria~\citep{miller1998noncircular, ConnorHastieTaylorProcRoySoc, CHT78, bishop198_equilibria}. Figure~\ref{fig:Greene-Chance-illustration} illustrates the main point --- we can change the gradient of the pressure and the safety factor locally by a finite amount without significantly changing their respective values.
 We will use this idea again in~\S\ref{sec:ITG-study} and~\S\ref{sec:TEM-study} to vary the pressure gradient at the nominal magnetic shear when we examine the microstability of different equilibria. Details explaining the Greene-Chance analysis are given in appendix~\ref{app:Greene-Chance-method}. 

To obtain useful maximum growth rate scans it is computationally advantageous to know where the equilibrium transitions from being stable to unstable, i.e., the region of marginal stability. This is because stable modes are extended and require many more points and a wider range in $\theta$ than unstable modes, leading to a longer computation time. To that end, we first integrate the marginally-stable ballooning equation 
\begin{equation}
    \frac{1}{\mathcal{J}}\frac{\partial}{\partial \theta}\left( \frac{\lvert \bnabla\alpha \rvert^2 }{\mathcal{J}\, B^2} \frac{\partial \hat{X}}{\partial \theta}\right) + 2 \frac{dp}{d\psi}\left[ \boldsymbol{B}\times\bnabla\left(p + \frac{B^2}{2}\right)\cdot\bnabla{\alpha}\right] \hat{X} = 0, 
    \label{eqn:marginal-ballooning-equation}
\end{equation}
along the field and count the zeros of the function $\hat{X}(\theta)$ --- if $\hat{X}$ has at least one zero, the mode is unstable, else it is stable. This criterion was originally developed by~\citet{newcombcriterion} for a screw pinch. He used it as a method to asses the stability of a screw pinch without explicitly finding the growth rates or the eigenfunctions. It is briefly explained in appendix~\ref{app:Newcomb's-criterion}. Using this criterion, one can obtain the sign of $\gamma^2$ and infer the stability significantly more rapidly than by exactly solving$~\eqref{eqn:ballooning-equation}$. Coupling Newcomb's criterion with the Greene-Chance analysis gives us the ability to scan the $\hat{s}-\alpha_{\mathrm{MHD}}$ space and plot the marginal stability contour ($\gamma = 0$) cheaply. For axisymmetric equilibria, the marginal stability contour is a single continuous line. Upon obtaining the contour, we choose a region around it where we solve $\eqref{eqn:ballooning-equation}$.

To solve~\eqref{eqn:ballooning-equation} we use the procedure described by Sanchez et al. \citep{sanchez2000cobra}. Our two-part code\footnote{Our \textrm{Python} code is freely available at \href{https://github.com/rahulgaur104/ideal-ballooning-solver}{github.com/rahulgaur104/ideal-ballooning-solver}.} first finds the contour of marginal stability, then takes a region around the contour in the $\hat{s}-\alpha_{\mathrm{MHD}}$ space and implements the algorithm given in~\citep{sanchez2000cobra}. It outputs the maximum eigenvalue  and the corresponding eigenfunction for each value of $\hat{s}$ and $\alpha_{\mathrm{MHD}}$. The plots of the maximum eigenvalue along with the curve of marginal stability are shown in the next section.\footnote{All the calculations in this work are done for $\theta_0 = 0$. A more complete picture would require one to scan over multiple values of $\theta_0$ and take the union of the resulting marginal stability curves and growth rate plots.}
\subsection{Ideal-ballooning analysis results}
\label{subsec:ideal-ballooning-results}
\begin{figure}
    \centering
    \includegraphics[width=\textwidth, trim={{0.021\textwidth} 0 {0.021\textwidth} {0.011\textwidth}}, clip]{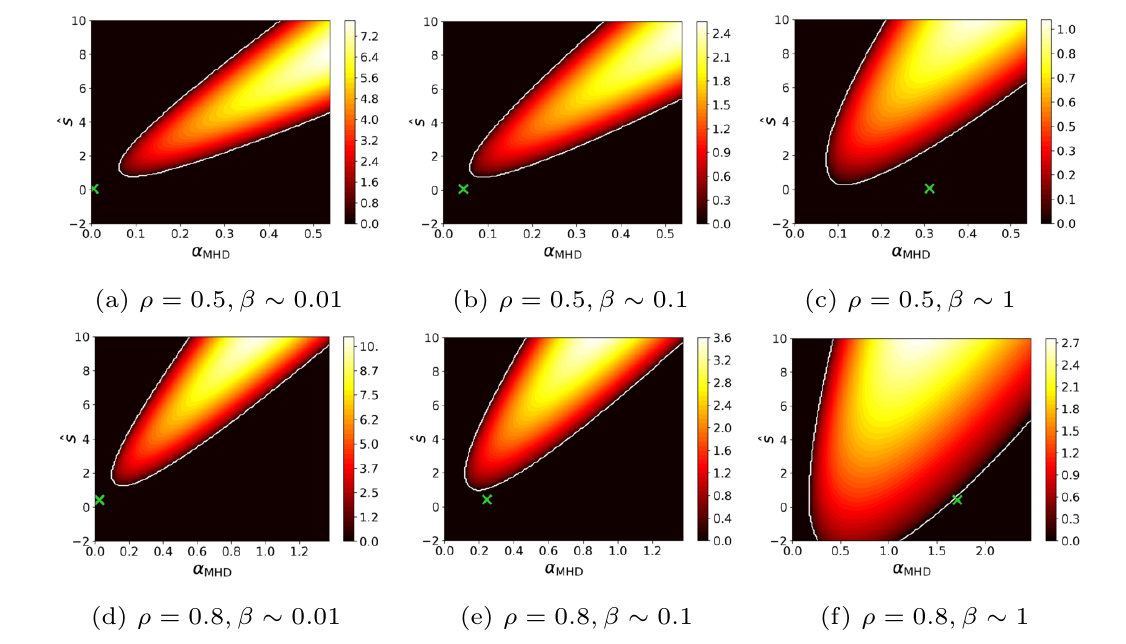}
\caption{This figure shows the normalized growth rate $\gamma a_{\rm{N}}/v_{\mathrm{th, i}}$ contours along with the curve of marginal stability (white line) for the positive triangularity equilibria. Columns correspond to the low, intermediate and high-$\beta$ regimes, respectively. The nominal equilibrium value is given by the green cross. Note the difference between the growth rates from the low and high-$\beta$ equilibria.}
\label{fig:postri-ballooning-study}
\end{figure}

This section contains the results of the $\hat{s}-\alpha_{\mathrm{MHD}}$ analyses of the twelve local equilibria that we chose in~\S\ref{subsec:VMEC}. We plot a 2D contour plot of the magnitude of the growth rate as defined by~\eqref{eqn:ideal-ballooning-growth-rate-defn}.  We begin by discussing the positive triangularity equilibria in figure~\ref{fig:postri-ballooning-study}.

All the positive triangularity equilibria studied here are stable at their nominal values. The low-$\beta$ equilibria lie below the marginal stability contour whereas the high-$\beta$ equilibria lie above it. For low-$\beta$ equilibria the ballooning threshold is well-known to be $\alpha_{\mathrm{MHD}} \sim 1$ but for the high-$\beta$ equilibria, using \eqref{eqn:alpha_MHD}, $\alpha_{\mathrm{MHD}} \sim 1/\delta_{\mathrm{Hsu}}^2 \gg 1$ which pushes these equilibria into the region of ``second'' stability, first discovered by~\citet{greene1981second}. Note that the high-$\beta$ equilibrium in figure 9(\textit{f}) is close to marginally stable. We also see that the maximum growth rate decreases as $\sqrt{\beta}$ --- high-$\beta$ equilibria having the smallest maximum growth rates. Next, we show this behavior with the ideal-ballooning scans of the negative triangularity equilibria.
\begin{figure}
    \centering
    \includegraphics[width=\textwidth, trim={{0.021\textwidth} 0 {0.021\textwidth} {0.011\textwidth}}, clip]{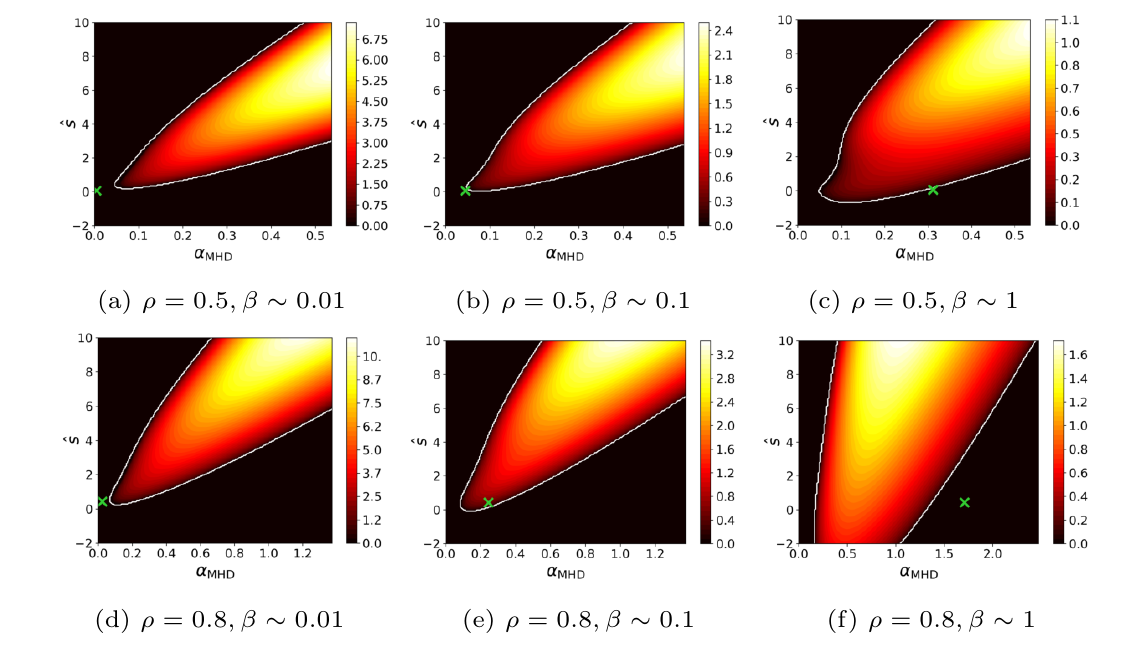}
\caption{This figure shows the normalized growth rate contours along with the curve of marginal stability for the negative triangularity equilibria. The nominal equilibrium value is denoted by the green cross.}
\label{fig:negtri_ball}
\end{figure}

In figure $\ref{fig:negtri_ball}$, we can see the nominal equilibria for the negative triangularity are stable for all cases except figure 10(\textit{e}). The trends follow those of the positive triangularity equilibria with two important exceptions. Firstly, unlike the positive triangularity equilibria, the high-$\beta$ negative triangularity equilibria get closer to the marginal stability line as we move towards the core. This is different from the expected result and might reveal certain advantages of negative triangularity geometry if explored further. Secondly, the growth rates in the unstable region in figure 10(\textit{e}) are lower than figure 10(\textit{f}). The most important takeaway from this study is that all low- and high-$\beta$ equilibria are stable to ideal-ballooning modes. This indicates that it might be possible to generate high-$\beta$ equilibria that are ballooning stable.\footnote{Note that we do not prove the experimental accessibility of these high-$\beta$ equilibria. We show that if these equilibria were to exist, they will be stable to the ideal-ballooning mode. The problem of accessibility in the context of ideal-ballooning stability was studied by~\citet{chance1990ideal}.} Even though one of the intermediate-$\beta$ equilibria is unstable, we will use it in our study as it will help us understand the behavior of microstability with changing $\beta$.     

\section{Microstability analysis}
\label{sec:microstability}
This section contains the general theoretical and numerical details of our microstability analysis. In~\S\ref{subsec:electrostatic-GK}, we will explain the physical basis and theoretical details of the gyrokinetic model. In the next section, we will explain how the model is implemented numerically using the \texttt{GS2} code and provide the general details of our numerical study.
\subsection{The gyrokinetic model}
\label{subsec:electrostatic-GK}
The electromagnetic gyrokinetic model is a simplification of the $6D$ Vlasov-Maxwell system of equations to a $5D$ system that predicts the self-consistent evolution of a distribution of charged particles and its electromagnetic fields in the presence of low-frequency, small-scale fluctuations. We define the distribution and fields as
\begin{equation}
\begin{gathered}
    f_s = F_{0s} + \delta f_s,\\
    \boldsymbol{E} = \boldsymbol{E}_0 + \delta \boldsymbol{E},\\
    \boldsymbol{B} = \boldsymbol{B}_0 + \delta \boldsymbol{B},\\
\end{gathered}
\end{equation}
where the fields comprise their equilibrium components (with a subscript $0$) plus small fluctuations. The fluctuations $\delta f_s, \delta \boldsymbol{E}$ and $\delta \boldsymbol{B}$ are defined such that they vanish when averaged over length and time scales much larger than the particle gyroradius $\rho_s$ and turbulence frequency $\omega$, respectively.
The gyrokinetic model is derived in the limit 
\begin{equation}
    \tilde{\epsilon} \equiv \frac{\omega}{\Omega_{s}} \sim  \frac{\rho_s}{a_{\rm{N}}} \sim \frac{k_{\parallel}}{k_{\perp}} \sim \frac{\delta f_s}{F_{0s}} \sim \frac{Z_s e \varphi}{T_{s}} \sim  \frac{|{\delta \boldsymbol{B}}|}{|\boldsymbol{B}_0|} \ll 1,
\end{equation}
where $\Omega_{s} = (Z_s e B)/(m_{s}c)$ is the cyclotron frequency and subscript $s$ denotes the species. The variable $\varphi$ is the perturbed electrostatic potential. The particle gyroradius $\rho_s$, given by
\begin{equation}
\rho_s \equiv \frac{v_{\mathrm{th},s}}{\Omega_s},
\end{equation}
is the perpendicular length scale of the turbulent fluctuations and $v_{\mathrm{th},s} = \sqrt{(2\,T_s)/m_s}$ is the thermal velocity. The length scale $a_{\mathrm{N}}$ is the effective minor radius, defined in~\S\ref{subsec:VMEC}. The wavenumbers of the mode perpendicular and parallel to the equilibrium magnetic field are denoted by $k_{\perp}$ and $k_{\parallel}$, respectively. 

In the small $\tilde{\epsilon}$ limit, one can reduce the dimensionality of the problem from $6D (\boldsymbol{r}, w_{\perp}, w_{\parallel}, \vartheta)$ to $5D (\boldsymbol{r}, w_{\perp}, w_{\parallel})$ by averaging over the gyrophase $\vartheta$. For this $5D$ coordinate system, we will transform back and forth between two different coordinates, the particle position and velocity coordinates $(\boldsymbol{r}, w_{\perp}, w_{\parallel}, t)$ and the guiding center coordinates $(\boldsymbol{R}_s, E_s, \mu_s, t)$ where
\begin{align}
    E_s &= \frac{1}{2} m_s w^2, \\
    \mu_s &= \frac{m_s w_{\perp}^2}{2B},
\end{align}
are the total kinetic energy and the magnetic moment of the particle. The guiding center is given in terms of the particle position by the Catto transformation~\citep{catto1977col}:
\begin{equation}
    \boldsymbol{R}_s = \boldsymbol{r} - \frac{\boldsymbol{b}\times \boldsymbol{w}_{\perp}}{\Omega_s}.
\end{equation}
The gyroaveraging operators $\langle\rangle_{\boldsymbol{R}_s}$ and $\langle\rangle_{\boldsymbol{r}}$
\begin{equation}
    \langle X \rangle_{\boldsymbol{R}_s} = \frac{1}{2\pi}\int_{0}^{2\pi} X(\boldsymbol{r}) d\vartheta,
\end{equation}
\begin{equation}
    \langle X \rangle_{\boldsymbol{r}} = \frac{1}{2\pi}\int_{0}^{2\pi} X(\boldsymbol{R}_s) d\vartheta,
\end{equation}
denote the average of $X$ over a gyration period at fixed guiding center $\boldsymbol{R}_s$ and at fixed position $\boldsymbol{r}$, respectively. It is convenient to define the gyrokinetic model in terms of the parallel component $\delta A_{\parallel}$ of the magnetic vector potential, the magnetic field strength fluctuation $\delta B_{\parallel}$
\begin{equation}
    \delta B_{\parallel} = \boldsymbol{b}\bcdot (\bnabla \times \delta \boldsymbol{A}_{\perp}),
\end{equation}
the electrostatic potential 
\begin{equation}
    \delta \boldsymbol{E} = -\bnabla\varphi - \frac{1}{c}\frac{\partial \boldsymbol{A}}{\partial t}, 
\end{equation}
and the gyrokinetic distribution function in the guiding-center coordinate system $ (\boldsymbol{R}_s, E_s, \mu_s, t)$
\begin{equation}
    h_s(\boldsymbol{R}_s, E_s, \mu_s, t) = -\frac{Z_s e \varphi(\boldsymbol{r}, t)  F_{0s}}{T_s} + \delta\!f_s(\boldsymbol{R}_s, E_s, \mu_s, t).
\end{equation}
Using these new fields we can now introduce the $\delta\!f$ gyrokinetic theory that was first derived for the linear electromagnetic case by~\citet{antonsen1980kinetic}  and non-linear case by~\citet{frieman1982nge}. For a collisionless, linear electromagnetic model, following the notation of~\citet{abel2013multiscale}, the governing equations are:
\begin{equation}
\frac{\partial h_s}{\partial t} + (w_{\parallel} \boldsymbol{b}+ \boldsymbol{v}_{Ds})\bcdot \frac{\partial h_s}{\partial \boldsymbol{R}_s} = \frac{Z_s e F_{0s}}{T_s}\frac{\partial \langle \varphi  - \boldsymbol{w} \bcdot \delta \boldsymbol{A}/c \rangle_{\boldsymbol{R}_s}}{\partial t}  + \boldsymbol{V}_{E} \bcdot \bnabla F_{0s},
\label{eqn:electrostatic-GK-equation}
\end{equation}
\begin{equation}
\sum_{s} \frac{(Z_s e)^2 \varphi}{T_s} = \sum_s Z_s e\int d^3\boldsymbol{w}\, \langle h_{s}\rangle_{\boldsymbol{r}}, \quad \tau = \frac{T_{\mathrm{e}}}{T_{\mathrm{i}}}\label{eqn:Poisson's-equation},
\end{equation}
\begin{equation}
    -\bnabla_{\perp}^2 \delta A_{\parallel} = \frac{4\pi}{c} \sum_s Z_s e \int d^3\boldsymbol{w}\, w_{\parallel} \langle h_s\rangle_{\boldsymbol{r}},
    \label{eqn:Parallel-Ampere's-Law}
\end{equation}
\begin{equation}
    \bnabla_{\perp}^2 \frac{\delta B_{\parallel}B}{4\pi} = -\bnabla_{\perp} \bnabla_{\perp} \boldsymbol{:}   \sum_s \int d^3\boldsymbol{w}\,  \langle  m_s \boldsymbol{w}_{\perp} \boldsymbol{w}_{\perp} h_s\rangle_{\boldsymbol{r}},
    \label{eqn:Perpendicular-Ampere's-Law}
\end{equation}
where the velocity integrals in $\eqref{eqn:Poisson's-equation}$ are taken at fixed $\boldsymbol{r}$. The velocities $\boldsymbol{V}_E$ and $\boldsymbol{v}_{Ds}$ are the $\boldsymbol{E}\times\boldsymbol{B}$ and the magnetic drift velocities, respectively:
\begin{align}
 \boldsymbol{V}_{E} &= \frac{c}{B} \boldsymbol{b}\times \langle \bnabla \varphi \rangle_{\boldsymbol{R}_s} - \frac{1}{B} \boldsymbol{b}\times \langle \bnabla( \boldsymbol{w}\bcdot\delta \boldsymbol{A}) \rangle_{\boldsymbol{R}_s},\\
\boldsymbol{v}_{Ds} &= \frac{w_{\parallel}^2}{\Omega_s} \boldsymbol{b}\times (\boldsymbol{b}\bcdot \bnabla\boldsymbol{b}) + \frac{w_{\perp}^2}{2\,\Omega_s} \frac{\boldsymbol{b}\times \bnabla B}{B}.
\label{eqn:particle-drift}
\end{align}
This completely defines the linear gyrokinetic system. The gyrokinetic model is a good approximation for the core region of a tokamak plasma where our study is being conducted~\citep{white2013multi, creely2017validation}. We solve the gyrokinetic model in a 3D flux tube ---  a tube with a rhombus-shaped cross section following the field line. The appropriate length of the flux tube and the boundary conditions at the ends are determined using the ideas developed by~\citet{beer1995field}. All the variables are assumed to be periodic perpendicular to the field line. This allows us to further simplify~\eqref{eqn:electrostatic-GK-equation} by writing the fluctuating fields as a Fourier series: 
\begin{equation}
\begin{gathered}
h_s = \sum_{k} h_{k_{\perp}, s}(\theta, t)\exp(i \boldsymbol{k}_{\perp}\bcdot \boldsymbol{R}_s ),\\
\varphi = \sum_{k} \varphi_{k_{\perp}}(\theta, t)\exp(i \boldsymbol{k}_{\perp}\bcdot \boldsymbol{r} ),\\
\delta A_{\parallel} = \sum_{k} \delta A_{\parallel,k_{\perp}}(\theta, t)\exp(i \boldsymbol{k}_{\perp}\bcdot \boldsymbol{r} ),\\
\delta B_{\parallel} = \sum_{k} \delta B_{\parallel, k_{\perp}}(\theta, t)\exp(i \boldsymbol{k}_{\perp}\bcdot \boldsymbol{r} ).
\label{eqn:normal-mode-ansatz}
\end{gathered}
\end{equation}
Applying this ansatz to~$\eqref{eqn:electrostatic-GK-equation}$, we obtain
\begin{equation}
\begin{split}
    \left(\frac{\partial}{\partial t} - i\omega_{Ds}\right)&h_{k_{\perp},s} + (\boldsymbol{b}\bcdot\bnabla\theta)  w_{\parallel} \frac{\partial h_{k_{\perp},s}}{\partial \theta } =  \left\{\frac{\partial}{\partial t} - i{\omega}_{*,s}\left[1 + \eta_s\left(\frac{E_s}{T_s} - \frac{3}{2}\right)\right]\right\} \times \,   \\ 
    &\Bigg[ J_0\left(\frac{k_{\perp}w_{\perp}}{\Omega_s}\right) \left(\varphi_{k_{\perp}} -\frac{w_{\parallel}\delta A_{\parallel}}{c}\right) + J_1\left(\frac{k_{\perp}w_{\perp}}{\Omega_s}\right) \frac{w_{\perp}}{k_{\perp}} \frac{\delta B_{\parallel}}{c}\Bigg] F_{0s},
\end{split}
\label{eqn:electrostatic-gyrokinetic-normal-mode}
\end{equation}
where  
\begin{equation}
    \omega_{Ds} = \boldsymbol{k}_{\perp}\bcdot \boldsymbol{v}_{Ds},
    \label{eqn:Particl-drift-frequency}
\end{equation}
is the magnetic drift frequency, $J_0(k_{\perp}\rho_s)$ and $J_1(k_{\perp}\rho_s)$ are the zeroth and first-order cylindrical Bessel functions, respectively, 
\begin{equation}
\frac{a_{\rm{N}}}{L_{\mathrm{T}s}} = - \frac{d\log(T_s)}{d\rho},\quad  \frac{a_{\rm{N}}}{L_{\mathrm{n}_s}} = - \frac{d\log(n_s)}{d\rho}, \quad \eta_s = \frac{L_{\mathrm{n}_s}}{L_{\mathrm{T}_s}}, 
\end{equation}
and
\begin{equation}
    \omega_{*,s} = \frac{T_s}{Z_s e B}\left[(\boldsymbol{b} \times \boldsymbol{k}_{\perp})\bcdot \bnabla \log{n_{s}}\right].
\end{equation}
For this study, we choose a Hydrogen plasma, i.e., $Z_{\mathrm{i}} = 1, Z_{\mathrm{e}}=-1$. The variables $L_{\mathrm{n}s}$ and $L_{\mathrm{T}s}$ are the density and temperature gradient scale lengths. The quantity $\omega_{*,s}$, the diamagnetic drift frequency, is the typical frequency of a drift wave --- waves caused due to gradients in temperature or density. For these modes
\begin{equation}
\frac{\partial}{\partial t} \sim  \omega \sim  \omega_{*,s}.
\end{equation}
In~\S\ref{sec:ITG-study} and~\S\ref{sec:TEM-study}, we will investigate two types of electrostatic drift wave instabilities, the Ion-Temperature Gradient (ITG) mode and the Trapped Electron Mode (TEM). For purely electrostatic modes, we solve equations~\eqref{eqn:electrostatic-GK-equation}-\eqref{eqn:Poisson's-equation} only and assume the magnetic fluctuations, $\delta A_{\parallel}$ and $\delta B_{\parallel}$ to be absent. In~\S\ref{sec:electromagnetic-GK}, we investigate the effect of magnetic fluctuations by solving the full set of equations~\eqref{eqn:electrostatic-GK-equation}-\eqref{eqn:Perpendicular-Ampere's-Law}. The procedure for numerically solving the gyrokinetic model is explained in the following section.

\subsection{Using the \texttt{GS2} code}
\texttt{GS2}\footnote{The \texttt{GS2} version used for this study is freely available at  \href{https://zenodo.org/record/4461680}{https://zenodo.org/record/4461680}.}~\citep{gs2ref, jenko2001nonlinear, dorlandETG, highcock2012thesis} is a parallel code that solves the gyrokinetic model as an initial-value problem. It solves equations~\eqref{eqn:electrostatic-GK-equation}-\eqref{eqn:Perpendicular-Ampere's-Law} numerically by calculating the evolution of an initial perturbation inside a flux tube.  

Before each run, one has to specify the value of the gradient scale lengths  $a_{\mathrm{N}}/L_{\mathrm{n_{\textit{s}}}}, a_{\mathrm{N}}/L_{\mathrm{T_{\textit{s}}}}$, the range of normalized wavenumbers $k_{\perp} \rho_{\rm{i}}$ and various geometric coefficients as a function of $\theta$. The exact values of these inputs depend on the mode under consideration and will be provided in the following sections.
Since we are only studying stability against fluctuations that vary on a small perpendicular scale $k_{\perp} \rho_{\rm{i}} \sim 1$, we can get the geometric coefficients from the local equilibria just like we did for the ideal-ballooning stability analysis.\footnote{Our \texttt{VMEC} to \texttt{GS2} interface for calculating the geometric coefficients is freely available at \href{https://github.com/rahulgaur104/VMEC2GK}{https://github.com/rahulgaur104/VMEC2GK}.} 

\begin{figure}
\centering
\includegraphics[width=0.9\textwidth, trim={0 {0.011\textwidth} 0 0}, clip]{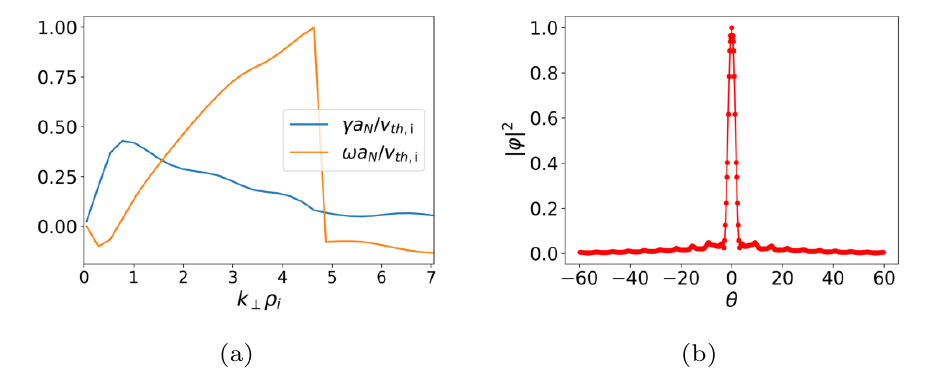}
\caption{This figure shows (\textit{a}) the output from a typical electrostatic \texttt{GS2} run showing the normalized frequency and growth rate spectrum and (\textit{b}) showing the variation along the field line of the square of the normalized electrostatic potential $\lvert\varphi\rvert^2$. We can see that the potential is well resolved and decays sufficiently before reaching the boundaries.}
\label{fig:GS2-typical-plots}
\end{figure}

The perpendicular structure of different fluctuations is defined by defining the wavevector $\boldsymbol{k}_{\perp}$ that can be written as
\begin{equation}
    \boldsymbol{k}_{\perp} = k_{y} \bnabla y + k_{x}\bnabla x.
\end{equation}
where $x$ and $y$ are normalized forms of the coordinates $\psi$ and $\alpha$, respectively.
For our study, we assume $k_{x} = 0$, i.e., modes with no variation in the radial direction.\footnote{This is equivalent to choosing $\theta_0 = 0$ in the ideal ballooning study. The parameter $\theta_0 \sim k_{x}/k_{y}$ denotes the tilt of a turbulent eddy with respect to the $\bnabla\psi$ direction. The most unstable modes almost always lie at $\theta_0 = 0$.} We choose around $15-25$ values of $k_y\rho_{\rm{i}}$ in the range $k_y\rho_{\rm{i}} = 0.05-6$. All of our simulations are well-resolved in $\theta$ and well-converged as can be seen in figure~\ref{fig:GS2-typical-plots}. For this study, $\theta \in [-10\pi, 10\pi]$ and more than $450$ points along the $\theta$ grid unless stated otherwise.

For the velocity space structure \texttt{GS2} uses an $(E, \lambda)$ grid instead of the $(w_{\parallel}, w_{\perp})$ grid. Defining the pitch angle as 
\begin{equation}
    \lambda(\theta_{\mathrm{b}}) = \frac{\mu}{E},
\end{equation}
where $\theta_{\mathrm{b}}$ is the bounce angle --- the value of $\theta$ at which a trapped particle with a pitch angle $\lambda$ reflects back from a region of high magnetic field. For a given pitch angle $\lambda$, the bounce angle is defined such that $B(\theta_b) = 1/\lambda$. In \texttt{GS2}, resolution of the passing particle distribution function in the coordinate $\lambda$ is set by the variable $\mathrm{nlambda}$. We set $\mathrm{nlambda} = 12$. For the trapped particle distribution, we choose $11$ bounce points. Similarly, for the energy-space resolution, we set the value of the \texttt{GS2} variable $\mathrm{negrid} = 10$.\footnote{For the sake of brevity, we have avoided explaining the details of the velocity grid. These details and resolution requirements can be found in much more detail in~\citet{highcock2012thesis}.} We choose an average of $27$ points along the flux tube for every $2\pi$ interval to ensure sufficient resolution along the field line.
This completely defines the resolution in \texttt{GS2}.

In figure~\ref{fig:GS2-typical-plots}, we show the results from a typical \texttt{GS2} run.
After each run, one obtains the normalized growth rate $\gamma a_{\mathrm{N}}/v_{\mathrm{th, i}}$, the wave frequency $\omega a_{\mathrm{N}}/v_{\mathrm{th, i}}$, the electrostatic potential eigenfunction $\varphi(\theta, t)$ for electrostatic runs and $\varphi(\theta, t), \delta A_{\parallel}(\theta, t)$ and $\delta B_{\parallel}(\theta, t)$ for electromagnetic runs. We also obtain the quasilinear particle and heat fluxes for each mode 
\begin{equation}
    \Gamma_{s, k_y} =  \int \frac{d\theta}{\boldsymbol{B}\bcdot\bnabla\theta} \int d^3\boldsymbol{w}\, (\boldsymbol{V}_{E, k_y}\bcdot\bnabla\psi)\, h_{s, k_y},
    \label{eqn:QL-flux1}
\end{equation}
\begin{equation}
    Q_{s, ky} = \int \frac{d\theta}{\boldsymbol{B}\bcdot\bnabla\theta} \int d^3\boldsymbol{w}\, (\boldsymbol{V}_{E, k_y}\bcdot\bnabla\psi) \, E_s\, h_{s, k_y},
    \label{eqn:QL-flux2}
\end{equation}
where the subscript $k_y$ denotes the mode $k_y$ in the Fourier space. These quantities can be used to extract information about an unstable mode~\citep{kotschenreuther2019fingerprints}. Since this is a linear study, the absolute values of the fluxes do not contain any useful information but their ratio $\Gamma_{s, ky}/Q_{s, ky}$  can still be used to characterize the type of instability. We will use this ratio for the TEM study in~$\S\ref{sec:TEM-study}$. 

\section{ITG study}
\label{sec:ITG-study}
This section contains the results and analysis of the ITG study of the equilibria that we obtained in~\S\ref{sec:equilibrium}. In~\S\ref{subsec:ITG-study-details}, we will present the specific details, including the values of different parameters used for the simulation and the reasoning behind our choices. In~\S\ref{subsec:local-shear}, we will introduce the local magnetic shear, a quantity that characterizes the stability of an equilibrium to the ITG mode. In the final section, we will plot and compare the results of all the different local equilibria and explain the stability of the high-$\beta$ equilibria. 
\subsection{Details of the study}
\label{subsec:ITG-study-details}
The most important form of electrostatic instability that arises at low wavenumbers, the ITG~\citep{cowley1991considerations}, occurs when a drift-wave becomes unstable due to a large ion temperature gradient, i.e., large $a_{\rm{N}}/L_{\mathrm{T_i}}$. Therefore, our objective is to understand this mode by doing a scan in the temperature gradient scale length, $a_{\mathrm{N}}/L_{\mathrm{T_i}}$. Using the definition of the pressure $p_s = n_s T_s$, we can write
\begin{equation}
    \frac{a_{\rm{N}}}{L_{p_s}} \equiv -\frac{d \log(p)}{d\rho} =  -\frac{d \log(T_s)}{d \rho} -  \frac{d\log(n_s)}{d \rho} = -a_{\rm{N}} \left(\frac{1}{L_{n_s}} + \frac{1}{L_{T_s}}\right).
    \label{eqn:beta-betaprime_consistency}
\end{equation}
Using the equation above along with $\eqref{eqn:alpha_MHD}$, we can write
\begin{equation}
    \alpha_{\mathrm{MHD}} =  \frac{\beta}{2}\, \sum_{s} \frac{a_{\mathrm{N}}}{L_{\mathrm{p}_s}}\, \frac{\rho q^2}{2 \mu_0 \epsilon}.
    \label{eqn:beta-betaprime-consistency}
\end{equation}
Furthermore, recall that we can vary the normalized pressure gradient $\alpha_{\mathrm{MHD}}$ for a local equilibrium using the idea of Greene and Chance without recalculating the global equilibrium. This gives us the ability to self-consistently
vary the temperature and density gradient scale lengths for a fixed $\beta$ as long as we recalculate the local equilibrium for the resulting value of $\alpha_{\mathrm{MHD}}$.\footnote{Maintaining self-consistency is crucial to all local analyses. Violating~\eqref{eqn:beta-betaprime_consistency} can give rise to specious, non-physical instabilities~\citep{zhu2020generalized}.} Table~\ref{tab:Table-3} contains the nominal density, temperature and pressure gradient scale lengths, denoted $a_{\rm{N}}/L^{\mathrm{nom}}_{\rm{n}_{\mathrm{i}}}$, $a_{\rm{N}}/L^{\mathrm{nom}}_{\rm{T}_{\mathrm{i}}}$ and $a_{\rm{N}}/L^{\mathrm{nom}}_{p_{\mathrm{i}}}$, respectively.
\begin{table}
\begin{center}
\begin{tabular}{ccccccc}
 $\rho$ & $a_{\rm{N}}/L^{\mathrm{nom}}_{\rm{T}_{\mathrm{i}}}$ & $a_{\rm{N}}/L^{\mathrm{nom}}_{\rm{n}_{\mathrm{i}}}$ & 
 $a_{\rm{N}}/L^{\mathrm{nom}}_{\rm{p}_{\mathrm{i}}}$ & 
 $a_{\rm{N}}/L^{\mathrm{nom}}_{\rm{T}_{\mathrm{e}}}$ &
 $a_{\rm{N}}/L^{\mathrm{nom}}_{\rm{n}_{\mathrm{e}}}$ &
 $a_{\rm{N}}/L^{\mathrm{nom}}_{\rm{p}_{\mathrm{e}}}$\\[3pt]
 $0.5$  & $0.59$  & $0.21$ & $0.80$ & $0.59$ & $0.21$ & $0.80$\\
 $0.8$  & $3.00$  & $1.09$ & $4.09$ & $3.09$ & $1.09$ & $4.09$ \\
\end{tabular}
\caption{Nominal gradient scale length values}
\label{tab:Table-nom-scale-length}
\end{center}
\end{table}
These are the values obtained from the original local equilibrium generated by \texttt{VMEC} and are exactly the same for all the different beta and triangularity values. For the ITG mode study, we define
\begin{equation}
    \mathrm{fac} = \frac{dP}{d\rho}\Big/\left(\frac{dP}{d\rho}\right)_{\mathrm{nom}},
    \label{eqn:bishop_fac}
\end{equation}
as the ratio of actual pressure to the nominal pressure. We choose $\mathrm{fac} = (0.5, 1, 2, 4, 8)$ times the nominal pressure gradient for $\rho = 0.5$ and $\mathrm{fac} = (0.5, 1, 2, 4)$ times the nominal pressure gradient for $\rho = 0.8$. For each pressure gradient, we choose two density gradient scale lengths --- the nominal and half of the nominal value from the local VMEC equilibrium while varying the temperature gradient scale length consistently for each gradient scale length. Tables~\ref{tab:Table-5} and~\ref{tab:Table-6} contain the resulting values.
\begin{table}
\begin{minipage}[t]{0.48\linewidth}\centering
\begin{tabular}{cccccc}
 $a_{\rm{N}}/L_{\mathrm{n_i}}$ & \multicolumn{5}{c}{$a_{\rm{N}}/L_{T_{\mathrm{i}}} $ } \\[3pt]
 $0.21$  & $0.19$  & $0.59$  &  $1.39$ & $2.98$ & $6.19$ \\
 $0.10$  & $0.29$  & $0.69$ & $1.49$ & $3.08$ & $6.29$ \\
\end{tabular}
\caption{Values of $a/L_{\rm{T_i}}$ at $\rho = 0.5$}
\label{tab:Table-5}
\end{minipage}
\begin{minipage}[t]{0.48\linewidth}\centering
\begin{tabular}{ccccc}
$a_{\rm{N}}/L_{\mathrm{n_i}}$  & \multicolumn{4}{c}{$a_{\rm{N}}/L_{T_{\mathrm{i}}}$} \\[3pt]
      $1.09$ & $0.95$  & $3.00$ & $7.10$ & $15.27$ \\
      $0.54$ & $1.50$  & $3.55$ & $7.64$ & $15.82$ \\
\end{tabular}
\caption{Values of $a/L_{\rm{T_i}}$ at at $\rho = 0.8$}
\label{tab:Table-6}
\end{minipage}
\end{table}

These $18$ values of various scale lengths are exactly the same for all the triangularities as well as for all the different beta values due to the way we have defined $\rho$.
From previous observations and studies, we know that the typical peak ITG growth rate lies around $k_y \rho_{\rm{i}} = 1$. To capture the maximum growth rate, we calculate the growth rates in the range $k_y \rho_{\rm{i}} \in [0.05, 2]$. For ITG, we have made the common assumption of adiabatic electrons to exclude the effect of kinetic electrons on the ITG mode and avoid other modes like the TEM. 
Mathematically, this means one assumes $h_{k,e} = 0$ when solving equation $\eqref{eqn:electrostatic-gyrokinetic-normal-mode}$ for the electrons.

Using the values in tables~\ref{tab:Table-5} and~\ref{tab:Table-6}, we run \texttt{GS2} in the electrostatic limit ($\delta A_{\parallel} = 0, \delta B_{\parallel} = 0$) and obtain the maximum normalized growth rate $\gamma a_{\rm{N}}/v_{\mathrm{th,i}}$ for each of the $108$ cases, $18$ for each beta and each triangularity value. The results showing comparison between different beta values, boundary shapes and normalized radii will be shown in~\S\ref{subsec:ITG-results}. 
We expect the equilibria to become more stable to the ITG mode as we increase $\beta$. This behavior is well-known~\citep{hirose2000finite, jarmen1998itg} in the literature for low and intermediate-$\beta$ equilibria. To try and explain this trend, in the next section we look as the local shear as a characteristic quantity that explains stabilization of the ITG mode with increasing $\beta$.

\subsection{Characterizing stability to the ITG mode}
\label{subsec:local-shear}
\begin{figure}
\centering
\includegraphics[width=0.98\textwidth]{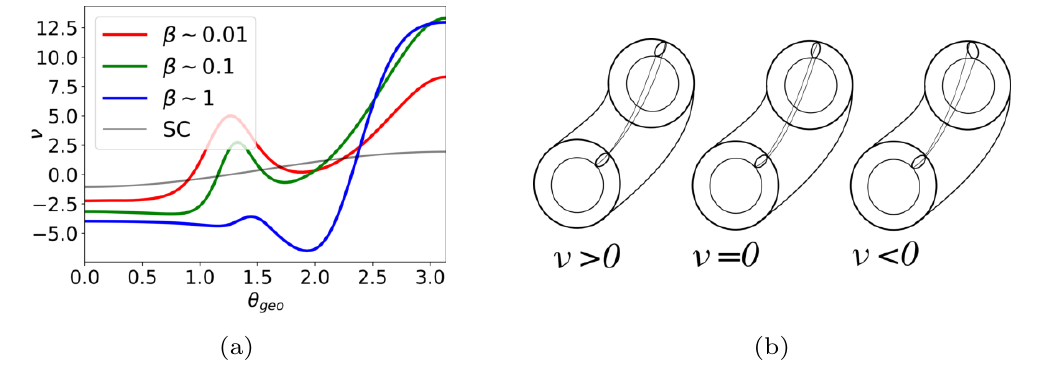}
\caption{This figure explains the physical meaning of the local magnetic shear with (\textit{a}) showing a typical local shear plot of negative triangularity equilibria at the nominal $\hat{s} = 0.45$ and nominal $\alpha_{\mathrm{MHD}}$ values at $\rho = 0.8$ for different beta values and a low-$\beta$ shifted circle model (abbreviated SC). On the right side (\textit{b}) illustrates a modified interpretation from~\citet{antonsen-drake1996physical} explaining the concept of local magnetic shear. Negative local shear twists the turbulent eddies more than positive or zero local shear, stabilizing the ITG mode.}
\label{fig:local-shear}
\end{figure}

In this subsection, we define and plot an important quantity, the local magnetic shear~\citep{greene1981second}, which will help us understand the response of a local equilibrium to the ITG mode. We use the local shear since negative global shear $\hat{s}$ is known to stabilize ITG~\citep{uchida2003stability}. In the following section, we will plot the local shear as a function of $\theta_{\mathrm{geo}}$ at the nominal $\hat{s}$ and $\alpha_{\mathrm{MHD}}$ for the high-$\beta$ equilibria and compare it with the low, intermediate and a low-$\beta$ shifted-circle equilibrium (abbreviated as SC in the plots).  We will show that the behavior of ITG is directly related to the local shear and argue that a large negative local shear over a wide range in $\theta_{\mathrm{geo}}$ stabilizes the ITG mode.
Mathematically, the local shear $\nu$ is given by~\citep{greene1981second, DewarGlasserballooning}
\begin{equation}
    \nu = -\boldsymbol{B}\cdot \bnabla_{\mathrm{N}}\left(\frac{\bnabla_{\mathrm{N}}\alpha\cdot {\bnabla}_{\mathrm{N}}\psi}{\lvert\bnabla_{N}\psi\rvert^2}\right),
\end{equation}
where $\bnabla_{\mathrm{N}} = a_{\mathrm{N}} \bnabla$ is used to non-dimensionalize $\nu$. To plot $\nu$ with respect to $\theta_{\mathrm{geo}}$, one needs to further simplify using the formalism given in appendix~\ref{app:Greene-Chance-method}. It is important to point out that
\begin{equation}
    \hat{s} = \frac{\rho |\bnabla_{\rm{N}} \psi|}{2\pi q\,  a^2_{\rm{N}}} \int_{0}^{2\pi} \frac{d\theta\, d\phi}{\boldsymbol{B}\cdot{\bnabla}_{\rm{N}}\theta}\,  \nu.
\end{equation}
Thus, the appropriately-weighted average of the local shear over a flux surface gives us the global magnetic shear $\hat{s}$.  The global shear is held fixed for all the local equilibria at a given $\rho$. This relation implies that the local shear can be negative for a given range in $\theta$ for a positive global shear $\hat{s}$. A typical plot of local shear at nominal $\hat{s}$ and $\alpha_{\mathrm{MHD}}$ is given in figure~\ref{fig:local-shear} for different beta values. 

\begin{figure}
    \centering
    \includegraphics[width=0.93\textwidth]{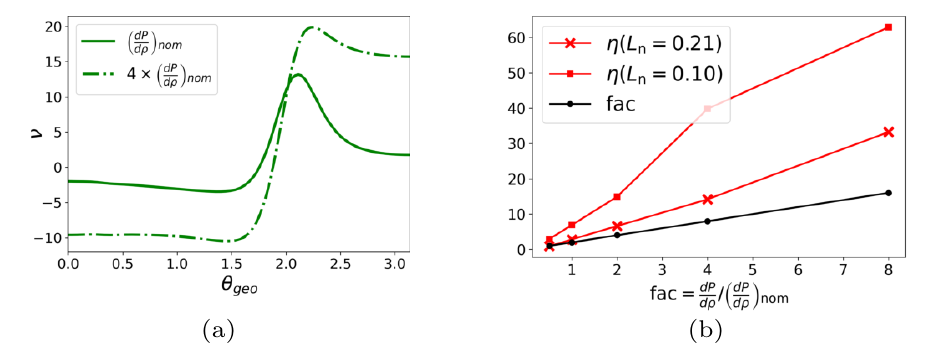}
    \caption{This shows (\textit{a}) the change in the local shear for $\mathrm{fac} = 4$ (increased pressure gradient). The local shear is much more negative with an approximately linear dependence with $\mathrm{fac}$ on outboard side. On the other hand, (\textit{b}) shows the comparison between the ITG driving term $\eta$ and the pressure gradient scaling factor $\mathrm{fac}$. The term $\eta$ is calculated using the values given in table~\ref{tab:Table-5}. We can see that $\eta$ grows linearly, but with a larger pre-factor. These figures illustrate how the ITG driving term grows more rapidly than the stabilizing local shear as we increase fac.}
    \label{fig:local-shear-fac}
\end{figure}

The local shear depends on the pressure gradient $\alpha_{\mathrm{MHD}}$ which further depends on $\beta$ as well as the gradient scale lengths
$L_{\mathrm{n}}$ and 
$L_{\mathrm{T}}$. The beta value increases the local negative shear through the Shafranov shift. The gradient scale length does so by increasing the poloidal current gradient $dF/d\rho$ (similar to~\eqref{eqn:lowest-order-Hsu}) required to balance the pressure gradient which consequently increases the toroidal magnetic field. A plot showing the effect of pressure gradient on the local shear is shown in figure~13\textit{(a)}. Note that decreasing the pressure gradient scale length  increases the driving term $\eta = L_{\mathrm{n}}/L_{\mathrm{T}}$. In fact, in this study, for a given $\beta$, $\eta$ increases faster than $a_{\mathrm{N}}/L_{\mathrm{p}}$. A plot of $\eta$ versus the pressure gradient scaling factor is shown in figure~\ref{fig:local-shear-fac}.
Hence, the dominant mechanism for generating negative local shear is the $\beta$-induced Shafranov shift. 
The local shear may also depend on the shaping, especially for negative triangularities. However, we think that shaping does not have a significant effect on the ITG stability of the high-$\beta$ equilibria.

\subsection{ITG results}
\label{subsec:ITG-results}
For convenience, we define a small positive ``threshold'' growth rate $\gamma_{\mathrm{thresh}}$ which we will use to separate stable and unstable modes.
If the maximum ITG growth $\mathrm{max}(\gamma a_{\mathrm{N}}/v_{th, \mathrm{i}}) < \gamma_{\mathrm{thresh}}$, we classify it as stable. 
For this study, we choose $\gamma_{\mathrm{thresh}} = 0.005$. We find that all the nominal equilibria stabilize as we increase $\beta$~\citep{candy2003egm} --- the high-$\beta$ equilibria are stable to the ITG mode, as shown in figure~\ref{fig:postri_nom_ITG} below. 
\begin{figure}
\centering
    \includegraphics[width=0.75\textwidth]{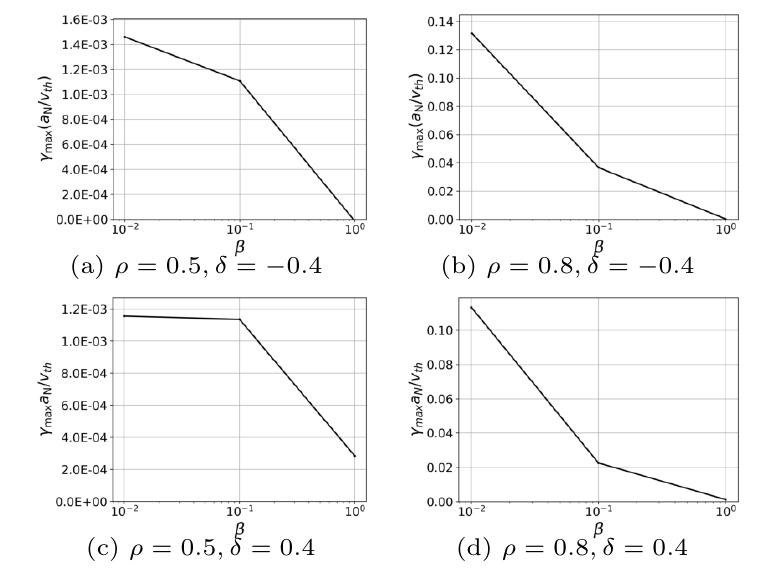}
\caption{These plots show the ITG max$(\gamma a_{\rm{N}}/v_{\mathrm{th, i}})$ (over $k_y \rho_{\rm{i}} \in [0.05, 2]$) vs the typical $\beta$ for nominal equilibria for negative triangularity case in the top row and positive triangularity case for the the bottom one. These plots are for the nominal equilibria only.}
\label{fig:postri_nom_ITG}
\end{figure}
To better understand what causes this effect, we do a scan in gradient scale length scales given in tables~\ref{tab:Table-5} and~\ref{tab:Table-6}. Each group of plots contains maximum growth rate for a range of $a/L_{\mathrm{T}_\mathrm{i}}$ at a fixed $a/L_{\mathrm{n}_\mathrm{i}}$ for the three beta values. Plots are grouped by the triangularity of the equilibria. For each group, there are subgroups based on the normalized radius $\rho$. The results corresponding to the positive triangularity boundary shape are shown in figure~\ref{fig:postri_ITG}. 
\begin{figure}
\centering
    \includegraphics[width=\textwidth]{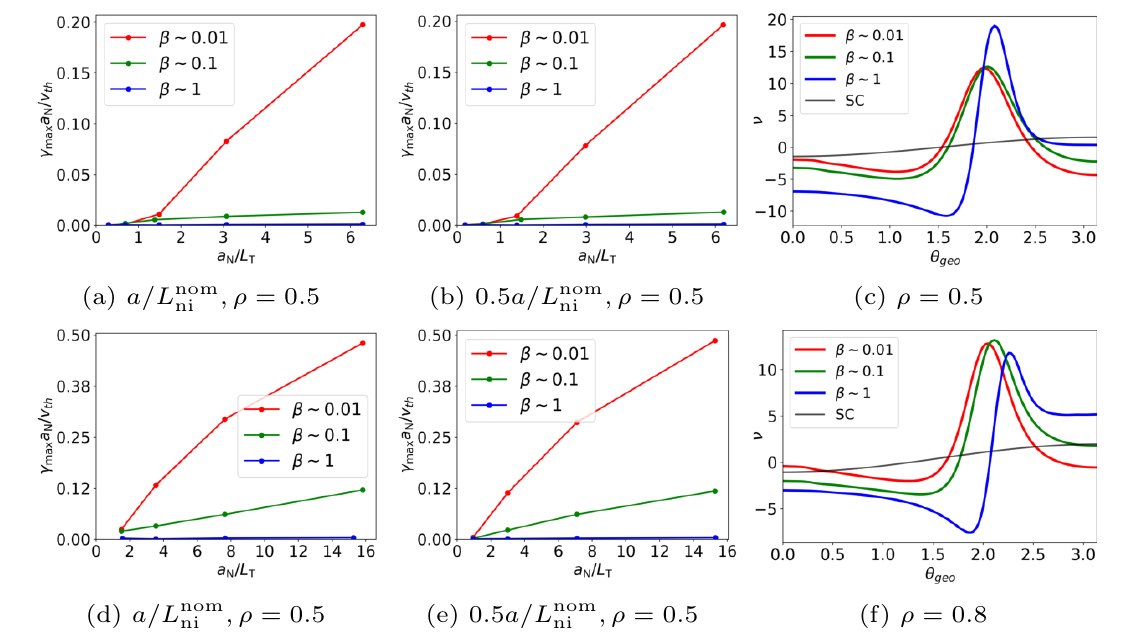}
\caption{shows the ITG max$(\gamma a_{\rm{N}}/v_{\mathrm{th, i}})$ plots for positive triangularity equilibria against the temperature gradient length scale. For the high-$\beta$ equilibria, ITG is stabilized at both $\rho = 0.5$ and $\rho = 0.8$. The right most figures in each row are the local magnetic shear vs the geometric theta $\theta_{\mathrm{geo}}$ at the nominal $dp/d\rho$ and $\hat{s}$. The grey line corresponds to the local shear for a shifted circle equilibrium (abbreviated SC). The magnetic shear $\hat{s}$ is the same for all the equilibria at  every $\rho$.}
\label{fig:postri_ITG}
\end{figure}

\begin{figure}
\centering
    \includegraphics[width=\textwidth]{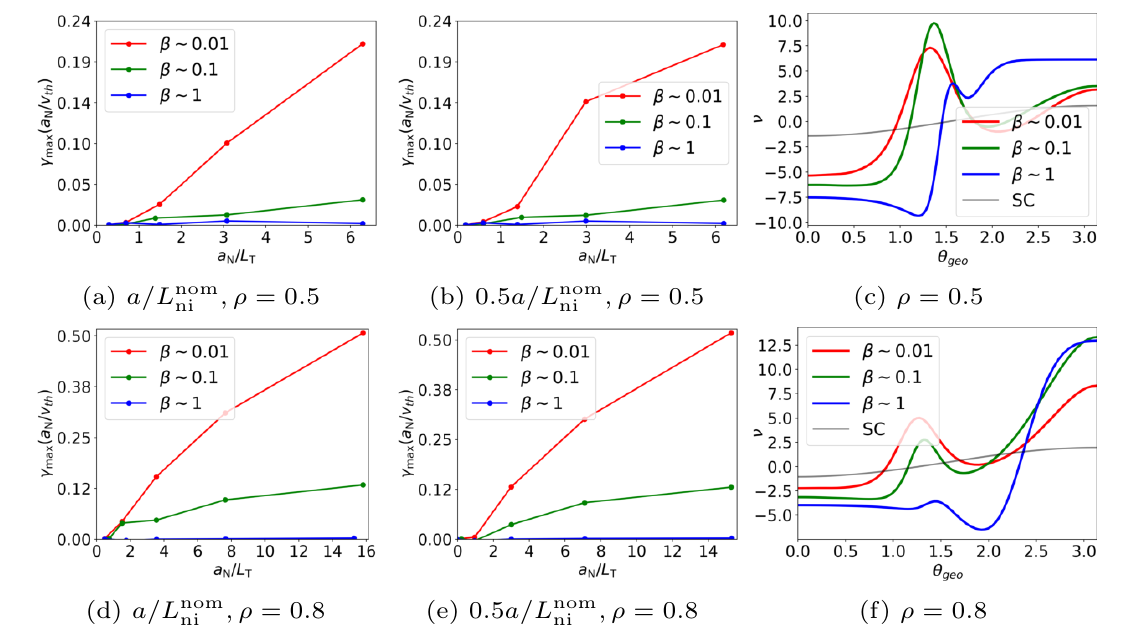}
\caption{shows the ITG max$(\gamma a_{\rm{N}}/v_{\mathrm{th, i}})$ plots for negative triangularity equilibria. For the high-$\beta$ equilibria, ITG is stabilized at both $\rho = 0.5$ and $\rho = 0.8$. The local shear for the high-$\beta$ equilibria is negative over the whole outboard side. The grey line corresponds to the local shear for a shifted circle equilibrium (abbreviated SC).}
\label{fig:negtri_ITG}
\end{figure}

In figure~\ref{fig:postri_nom_ITG}, one observes that for the high-$\beta$ cases, the ITG mode is stable (that is, $\gamma < \gamma_\mathrm{thresh})$. For the intermediate and low-$\beta$ cases, figure~\ref{fig:postri_ITG} shows destabilization with increasing temperature gradient. We believe that the stabilization of high-$\beta$ equilibria is a result of large local negative shear (right most panels) that spans over a wide range in $\theta$. The local shear becomes positive only after $\theta_{\mathrm{geo}} > \pi/2$ --- the whole outboard side has a large local negative shear. These large negative values are predominantly due to a large Shafranov shift but could have sub-dominant effects resulting from strong shaping, especially for the negative triangularity high-$\beta$ equilibria. For the shifted-circle equilibria, the local shear is small as compared with the rest of the equilibria even though it is negative over the outboard side. This is because the shifted circle equilibria are neither strongly shaped nor have a large beta. The trend remains the same even for the equilibria with half-nominal density gradients.
%
%
Next, we plot results for negative triangularity equilibria in figure~\ref{fig:negtri_ITG}.

We see that the high-$\beta$ equilibria with a negative triangularity boundary shape are stabilized as well. The local magnetic shear shows the same trend with beta but is even more negative and spans an even wider range in $\theta_{\mathrm{geo}}$ as compared to the positive triangularity equilibria. We believe this extended range in negative local shear is due to the ``squareness'' of the high-$\beta$ profiles, which is a stronger shaping than the positive triangularity cases. Since all the high-$\beta$ equilibria are stabilized and the growth rates are small, it is hard to find a clear difference based on triangularity. 

In the next section, we will study the stability of different equilibria to an electron-driven electrostatic mode, the Trapped Electron Mode (TEM).

\section{TEM study}
\label{sec:TEM-study}
In this section, we will present our analysis of the TEM instability. In~\S\ref{subsec:TEM-study-details}, we explain the TEM and tabulate the parameters for which we perform our study. After that, we define the maximum TEM growth rate. In~\S\ref{subsec:precession-drift}, we present the electron precession drift as a characteristic of TEM stability of a local equilibrium. In the final section, we plot the results in a fashion similar to the previous section.
\subsection{Details of the study}
\label{subsec:TEM-study-details}
The second type of electrostatic instability that we investigate, the collisionless TEM, becomes unstable when drift waves resonate with the precession of the electrons. This can cause significant electron loss that degrades plasma confinement~\citep{adam1976destabilization}. For the TEM, we choose four values of pressure gradient, corresponding to $\mathrm{fac} = (0.5, 1, 2, 4, 8)$ for $\rho = 0.5$ and five values corresponding to $\mathrm{fac} = (0.5, 0.75, 1, 2, 4)$ for $\rho = 0.8$. For each pressure gradient, we choose two temperature gradient scale lengths
--- nominal and $30\%$ of the nominal, while scanning the growth rates in the density gradient scale lengths. 
We do this since TEM, unlike ITG, is primarily a density gradient driven instability. The tables~\ref{tab:Table-7} and~\ref{tab:Table-8} contain the values for which we have solved the gyrokinetic model. Note that the temperature and density scale lengths are the same for the ions and electrons.
\begin{table}
\centering
\begin{tabular}{cccccc}
 $a_{\rm{N}}/L_{\mathrm{T_e}} = a_{\rm{N}}/L_{\mathrm{T_i}}$ & \multicolumn{5}{c}{$a_{\rm{N}}/L_{{\mathrm{n_e}}} = a_{\rm{N}}/L_{{\mathrm{n_i}}} $ } \\[3pt]
 $0.186$  & $0.21$  & $0.61$ & $1.41$ & $3.01$ & $6.22$ \\
 $0.296$  & $0.10$  & $0.51$  &  $1.31$ & $2.91$ & $6.11$ \\
\end{tabular}
\caption{Values of gradient scale lengths at $\rho = 0.5$}
\label{tab:Table-7}
\end{table}
\begin{table}
\centering
\begin{tabular}{cccccc}
$a_{\rm{N}}/L_{\mathrm{T_e}} = a_{\rm{N}}/L_{\mathrm{T_i}}$  & \multicolumn{5}{c}{$a_{\rm{N}}/L_{{\mathrm{n_e}}} = a_{\rm{N}}/L_{{\mathrm{n_i}}}$} \\[3pt]
$0.95$ & $1.09$ & $2.11$  & $3.13$ & $7.22$ & $15.45$ \\
$1.50$ & $0.54$ & $1.57$  & $2.59$ & $6.68$ & $14.86$ \\
\end{tabular}
\caption{Values of gradient scale lengths at at $\rho = 0.8$}
\label{tab:Table-8}
\end{table}

These $20$ values of gradients are the same for ions and electrons as well as both positive and negative triangularity equilibria at all the different beta values. Unlike the ITG study, we turn on the kinetic effects of electrons since TEM is an electron driven instability. The TEM growth rate peak occurs over a wide range $k_y\rho_{\rm{i}} \in [0.5, 6]$. Since there is an overlap with the ITG and the Electron-Temperature Gradient (ETG) mode, having two species makes it difficult to separate modes with purely ITG and ETG-related effects from modes with purely TEM-related effects. Therefore, to calculate the TEM growth rate, we choose the growth rate corresponding to the wavenumber $k_y\rho_{i}$ at which the ratio of the quasilinear electron flux to electron heat flux is the maximum, i.e., 
\begin{equation}
    \gamma_{\mathrm{TEM}}(k_y\rho_{\mathrm{i}}) = \gamma\Big \lvert_{\mathrm{max}(\Gamma_{\mathrm{e}, ky}/Q_{\mathrm{e}, ky})},
\end{equation}
where the definition of the quasilinear fluxes is given in equations~$\eqref{eqn:QL-flux1}$ and~$\eqref{eqn:QL-flux2}$. 

We run \texttt{GS2} for the wavenumbers in the range $k_y\rho_{\mathrm{i}} \in [0.2, 6.5]$. Just like the ITG study, we run \texttt{GS2} and obtain the maximum growth rate $\gamma a_{\rm{N}}/v_{\mathrm{th, i}}$ for each of the $120$ cases. The results showing comparison between different beta values, triangularities and the normalized radius will be shown in~\S\ref{subsec:TEM-results}.

Curvature driven TEMs are associated with the precession of trapped electrons in a flux surface. To that end, we elucidate the definition and role of the electron precession drift frequency in the next section and how it characterizes the stability of an equilibrium to the TEM.  

\subsection{Characterizing stability to the TEM}
\label{subsec:precession-drift}
\begin{figure}
\centering
\includegraphics[width=0.9\textwidth]{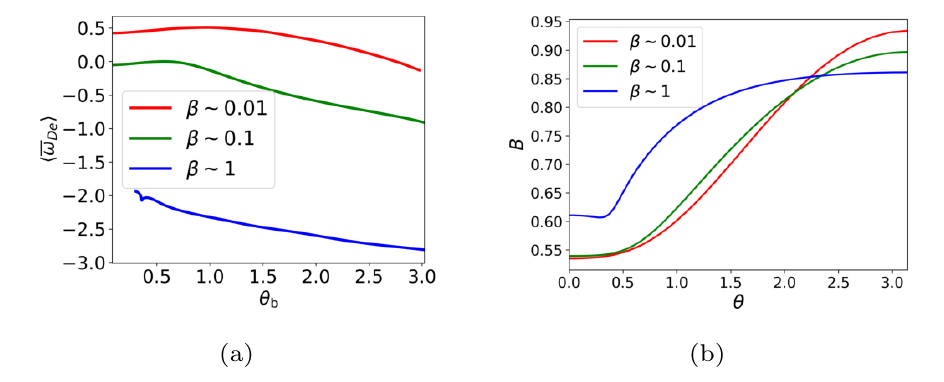}
\caption{This figure shows the precession drift in (\textit{a}) and the corresponding magnetic field magnitude in (\textit{b}) for negative triangularity equilibria at $\hat{s} = 0.45$ and nominal $\alpha_{\mathrm{MHD}}$ values at $\rho=0.8$ for different beta values. Note the atypical magnetic field for the high-$\beta$ equilibria where $\mathrm{min}(B)$ is located at a finite $\theta$.}
\label{fig:TEM-FOMS}
\end{figure}

The collisionless, curvature-driven TEM is a drift wave that becomes unstable when it resonates with the bounce precession of trapped electrons. The precession of electrons is characterized by their precession frequency
\begin{equation}
    \langle\omega_{\mathrm{De}}\rangle =  \left(\int_{-\theta_b}^{\theta_{\mathrm{b}}} \dfrac{d\theta}{w_{\parallel}} \frac{B}{(\boldsymbol{B}\bcdot {\bnabla}\theta)}\right)^{-1}  \int_{-\theta_b}^{\theta_{\mathrm{b}}} \frac{d\theta}{w_{\parallel}} \dfrac{B}{(\boldsymbol{B}\cdot {\bnabla}\theta)}\,  \omega_{\mathrm{De}},
\end{equation}
where the integral operator is the bounce-average operator. The precession frequency is a function of the bounce angle $\theta_{\mathrm{b}}$. Depending on the convention, one usually takes $\mathrm{sign}(\omega) = \mathrm{sign}(\omega_{*,\mathrm{e}})$ for trapped electron modes since the TEM is a drift wave. Therefore, if 
\begin{equation}
 \mathrm{sign}(\langle \omega_{\mathrm{De}} \rangle) \mathrm{sign}(\omega_{*,\mathrm{e}}) < 0, 
 \label{eqn:TEM-stability-condition}
\end{equation}
the drift wave will not be able to resonate with the precession of electrons. If the precession drift satisfies $\eqref{eqn:TEM-stability-condition}$ at all the different pitch angles, the curvature-driven TEM will be stabilized~\citep{connorTEM}.  
The expression for $\omega_{\mathrm{D}}$, given in $\eqref{eqn:Particl-drift-frequency}$ can be alternatively written as
\begin{equation}
    \omega_{\mathrm{De}} = \frac{k_y \rho_{\rm{e}}}{2} \frac{v_{\rm{th,e}}}{a_{\rm{N}}} E_{\rm{e}} \left[2(1-\lambda B) \omega_{\kappa} + \lambda B \omega_B \right]
\end{equation}
where $\omega_{\kappa}$ and $\omega_B$ are geometric factors independent of the electron energy $E_{\rm{e}}$ and the pitch angle $\lambda$. As a characteristic of TEM stability, we define the quantity
\begin{equation}
\bar{\omega}_{\mathrm{De}} = \langle \omega_{\rm{De}} \rangle \mathrm{sign}(\omega_{*,\mathrm{e}})/E_{\mathrm{e}},   
\end{equation}
as the precession drift per energy in the electron diamagnetic direction. A typical plot of the $\bar{\omega}_{\rm{De}}$ is shown at nominal $\hat{s}$ and $\alpha_{\mathrm{MHD}}$ for different $\beta$ values in figure~\ref{fig:TEM-FOMS}.
We see that the precession drift is negative everywhere only for high-$\beta$ case. This will form the basis for our understanding of the growth rate trends in the following section. 

\subsection{TEM results}
\label{subsec:TEM-results}
\begin{figure}
    \centering
\includegraphics[width=0.75\textwidth]{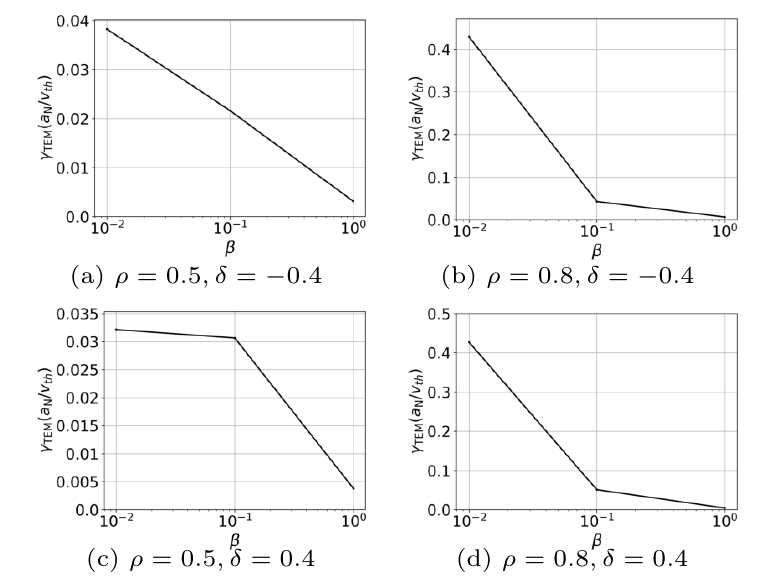}
\caption{shows the TEM $\gamma_{\mathrm{TEM}} (a_{\mathrm{N}}/v_{\mathrm{th, i}})$ vs the typical $\beta$ plots for nominal equilibria for negative triangularity case in the top row and positive triangularity case for the the bottom one.}
\label{fig:postri_nom_TEM}
\end{figure}

Just like the ITG study, we choose a value of a growth rate $\gamma_{\mathrm{thresh}}$ such that 
if $\gamma_{\mathrm{TEM}} (a_{\mathrm{N}}/v_{\mathrm{th, i}}) < \gamma_{\mathrm{thresh}}$,
we classify an equilibrium as stable. For this study, we choose $\gamma_{\mathrm{thresh}} = 0.005$.
First, we plot the maximum TEM growth rates for the nominal equilibria in figure~\ref{fig:postri_nom_TEM}. We find that increasing the beta stabilizes the TEM and the high-$\beta$ equilibria are stable to the TEM.

\begin{figure}
    \centering
    \includegraphics[width=\textwidth]{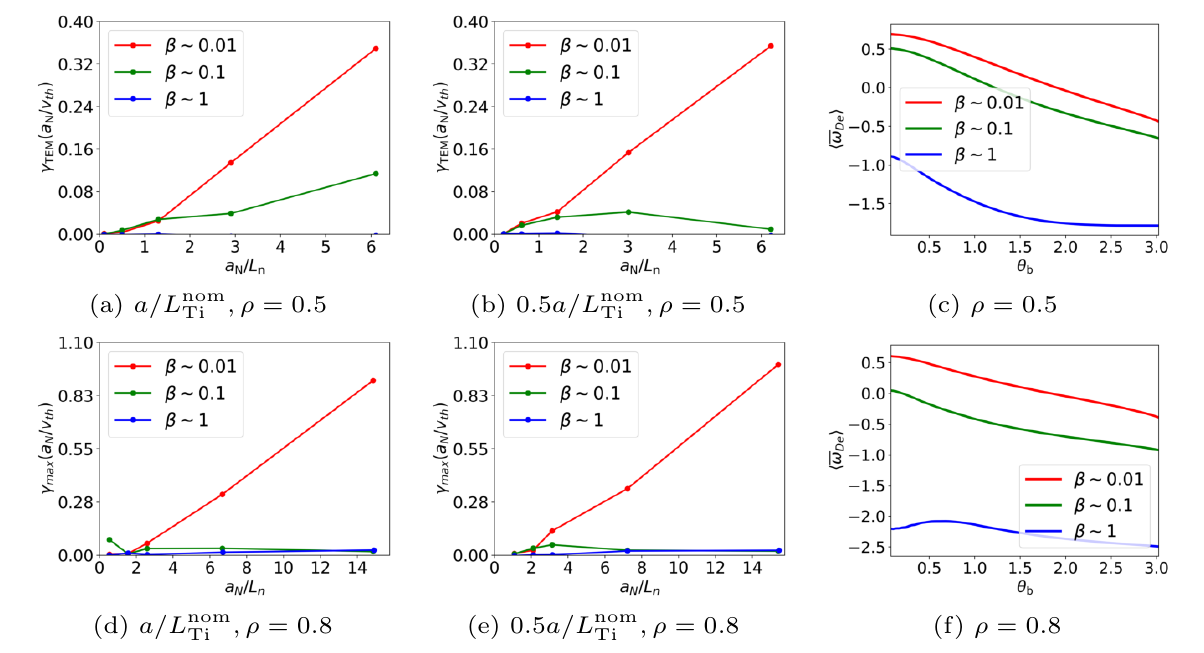}
\caption{shows the TEM $\gamma_{\mathrm{TEM}} (a_{\mathrm{N}}/v_{\mathrm{th, i}})$ plots for positive triangularity equilibria. For the high-$\beta$ equilibria, TEM is stabilized at both $\rho = 0.5$ and $\rho = 0.8$. As you can see in figures 19\textit{(c)} and 19\textit{(f)}, the precession drift for the high-$\beta$ equilibria is negative for all values of the bounce angle $\theta_{\rm{b}}$.}
\label{fig:postri_TEM}
\end{figure}

To understand this effect further, we plot the result from scans in the density gradients
in two groups, each group containing the maximum TEM growth rate for a range of $a/L_{\mathrm{n}}$ at a fixed $a/L_{\mathrm{T}}$, given in tables~\ref{tab:Table-7} and~\ref{tab:Table-8} for different equilibria. Just like the ITG study, we group the plots by triangularity and arrange them in rows based on the normalized radius $\rho$. For positive triangularity equilibria, the results are shown in figure~\ref{fig:postri_TEM}

We observe that the TEM is completely suppressed for the high-$\beta$, positive triangularity equilibria. The frequency $\bar{\omega}_{\mathrm{De}}$ is negative everywhere which means that all the trapped electrons precess in a direction opposite to the electron diamagnetic direction. They cannot destabilize the drift wave by exchanging energy with them.

The TEMs we seek are curvature driven. At higher densities, there exists another branch of instability that is slab-like and purely driven by the density gradient, called the universal mode~\citep{landreman2015universal}. We can see that the universal mode starts to dominate for the intermediate equilibria at large density gradients. For the high-$\beta$ equilibria, the universal mode is also suppressed since the large local shear, combined with strong shaping reduces the shearing length scale
\begin{equation}
    L_{\nu} = \frac{a_{\mathrm{N}}}{\nu}
\end{equation}
which makes it harder for fluctuations to grow and persist along the field line.

\begin{figure}
    \centering
    \includegraphics[width=\textwidth]{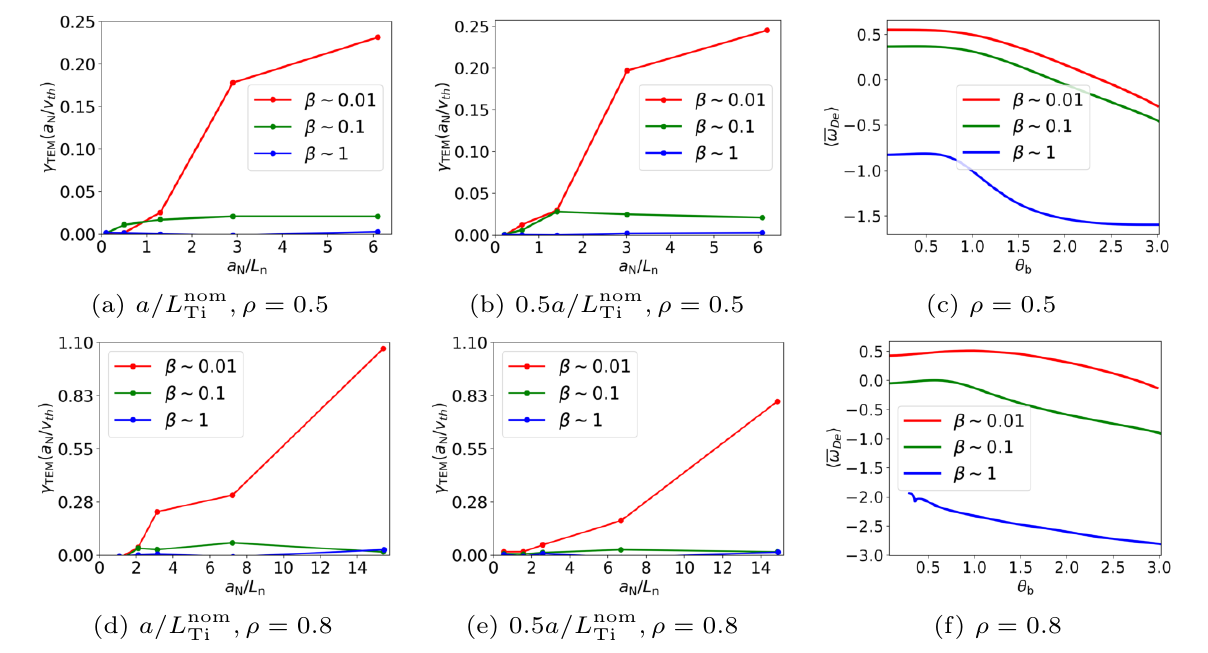}
\caption{shows the TEM $\gamma_{\mathrm{TEM}} (a_{\mathrm{N}}/v_{\mathrm{th, i}})$ plots for negative triangularity equilibria. For the high-$\beta$ equilibria, TEM is stabilized at both $\rho = 0.5$ and $\rho = 0.8$. The right most plots on each row is the electron precession drift frequency.}
\label{fig:negtri_TEM}
\end{figure}

Next, now look at the TEM growth rate trends for the negative triangularity equilibria, shown in figure~\ref{fig:negtri_TEM}.
The negative triangularity TEM growth rates follow the same trend as the positive triangularity ones. The TEM is suppressed for the high-beta equilibria due to the negative precession drift. The intermediate and low beta are more unstable for the negative triangularity equilibria at $\rho=0.5$ and as unstable as the positive triangularity ones at $\rho=0.8$. 

We have demonstrated the stability of high-$\beta$ equilibria against two major sources or electrostatic instabilities. However, when $\beta \sim 1$, magnetic fluctuations may play an important role in deciding the stability of an equilibrium. Therefore, in the next section, we study the effect of electromagnetic modes on the high-$\beta$ equilibria.   

\section{Linear Electromagnetic study}
\label{sec:electromagnetic-GK}
\begin{figure}
    \centering
    \includegraphics[width=\textwidth]{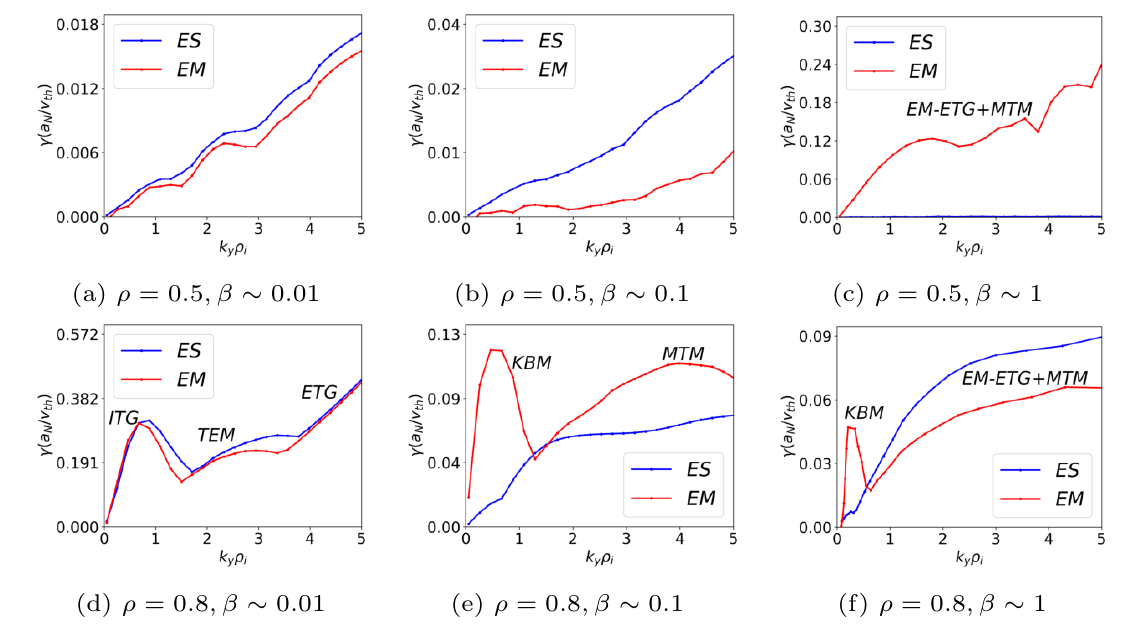}
\caption{This figure shows comparison between the electrostatic (abbreviated ES) adn electromagnetic (abbreviated EM) growth rates for all the nominal positive triangularity equilibria. Some of the branches have been labeled by their corresponding mode names. Notice the KBM in the intermediate and high-$\beta$ equilibria in figures 21(\textit{e}) and 21(\textit{f}), respectively and the emergence of the collisionless-micro-tearing and electromagnetic-ETG modes in figure 21(\textit{c}) for all values of $k_y\rho_{\rm{i}}$ and figures 21(\textit{e}) and 21(\textit{f}) for $k_y\rho_{\rm{i}} > 1$}
\label{fig:postri-EM-study}
\end{figure}

To see if the stability trend seen in the electrostatic study holds when we include electromagnetic effects, we perform an electromagnetic microstability analysis for all the the nominal local equilibria. We solve the linear, collisionless, gyrokinetic model allowing for non-zero magnetic field perturbations $\delta A_{\parallel}$ and $\delta B_{\parallel}$ using the \texttt{GS2} code. We use the nominal gradient scale lengths for this study (given in table~\ref{tab:Table-nom-scale-length}). First, we plot the growth rate spectrum with $k_y \rho_{\mathrm{i}}$ for the positive triangularity cases in figure~\ref{fig:postri-EM-study}.

Since \texttt{GS2} only calculates the maximum growth rate, in the inner core, we observe the finite-$\beta$ stabilization of ITG~\citep{candy2003eulerian} until the emergence of collisionless-microtearing (MTM)~\citep{applegate2007micro, guttenfelder2011electromagnetic, kotschenreuther1995quantitative} and electromagnetic-ETG (EM-ETG) modes~\citep{kim1991electromagnetic, joiner2007effects} in figure 21(\textit{c}). These modes arise when 
\begin{equation}
    k_y \rho_{\mathrm{e}} \sim k_y d_e \sim 1,
\end{equation}
where $\omega_{\mathrm{pe}}$ is the electron plasma frequency and $d_e = c/\omega_{\mathrm{pe}}$ is the electron skin depth. Since $\rho_{\mathrm{e}}/d_{\mathrm{e}} \sim \sqrt{\beta_\mathrm{e}}$, for low-$\beta$ equilibria $\rho_{\mathrm{e}} \gg d_{\mathrm{e}}$ whereas for high-$\beta$ equilibria $\rho_{\mathrm{e}} \sim d_{\mathrm{e}}$. Since these modes arise on small scales ($\textit{O}(\rho_{\rm{e}})$), they posses a fine radial structure and are therefore extended in the ballooning angle $\theta$. For the equilibria that are unstable to the EM-ETG and MTMs, we choose $\theta \in [-30\pi, 30\pi]$ with an average of $20$ points over a $2\pi$ interval to fully capture the fine radial structure.  

In the outer core, we observe finite-$\beta$ stabilization only for the low-$\beta$ equilibrium. For the intermediate and high-$\beta$ cases, the electrostatic modes are replaced by KBMs at low wavenumbers and collisionless-MTM and electromagnetic-ETG at high wavenumbers.
Overall, positive triangularity high-$\beta$ equilibria are more unstable than the low or intermediate ones in the inner core due to the collisionless-MTM and the EM-ETG mode. As we move towards the outer core, high-$\beta$ equilibria become much more stable --- exactly the opposite trend compared to the the rest of the equilibria. This means that the outer core is more stable for the high-$\beta$ equilibria.

\begin{figure}
    \centering
    \includegraphics[width=\textwidth]{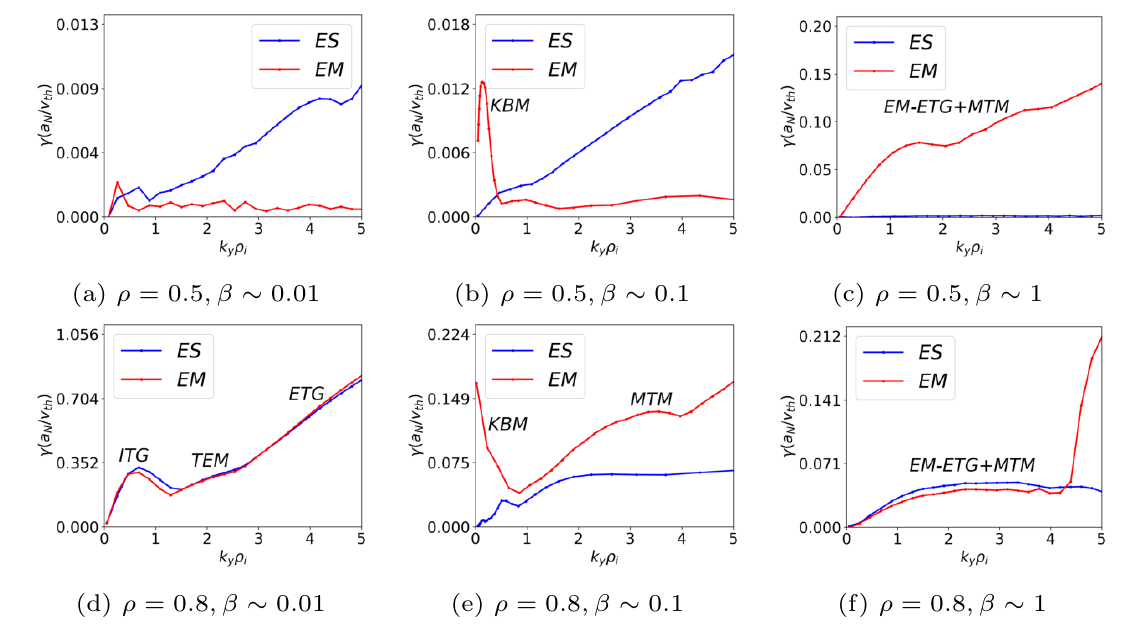}
\caption{This figure shows comparison between the electrostatic (ES) electromagnetic (EM) growth rates for all the nominal negative triangularity equilibria. The sudden jump in figure 22(\textit{f}) around $k_y\rho_{\mathrm{i}} = 4.5$ is a different branch of the collisionless MTM. Note also that the growth rate around $k_y \rho_{\mathrm{i}} = 0$ in figure 22(\textit{e}) does not go to zero since the equilibrium is unstable to the ideal ballooning mode.}
\label{fig:EM-negtri-plots}
\end{figure}

Next, we plot the growth rates for the negative triangularity equilibria in figure~\ref{fig:EM-negtri-plots}.
The negative triangularity, inner-core equilibria are also stabilized due to finite-$\beta$ effects. This effect is also visible for the intermediate-$\beta$ case for $k_y\rho_{\mathrm{i}} > 0.5$ but absent for the high-$\beta$ cases as the ITG and TEM are superseded by electromagnetic modes: collisionless-MTM and the EM-ETG mode. For the outer-core cases, we see the exact same pattern as the inner core --- finite-$\beta$ stabilization for the low-$\beta$, KBM and collisionles-MTMs for the intermediate-$\beta$ and collisionless-MTMs and EM-ETG modes for the high-$\beta$. Note that unlike the postive triangularity case, we do not observe the KBM in the negative triangularity, high-$\beta$ equilibria. Moreover, the growth rates are also smaller than the positive triangularity cases for a wide range of wavenumbers ($k_y \rho_{\mathrm{i}} \in [0.01, 4.5]$). Since turbulence is most likely to peak at lower wavenumbers, the growth rate characteristics of negative triangularity, high-$\beta$ are the most favorable.~\footnote{Some of the $\beta \sim 1$ equilibria have small but finite TEM growth rates which are larger than the growth rates from our electrostatic TEM study. This is consistent with the electrostatic study since we chose the TEM growth rate at a value of $k_y \rho_{\mathrm{i}}$ for which the ratio of $\Gamma_e/Q$ was maximized. This almost always happens at a low $k_y\rho_{\mathrm{i}}$ --- TEM growth rates are small at low $k_y\rho_{\mathrm{i}}$ values for $\beta \sim 1$. Moreover, the values of the gradient scale lengths (tables~\ref{tab:Table-7} and~\ref{tab:Table-8}) were different for the electrostatic study.}

In summary, we find that turning on the electromagnetic effects destabilizes the high-$\beta$ equilibria. Figures 21(\textit{c}), 22(\textit{c}) and 22(\textit{f})
show the emergence of collisionless-MTMs and EM-ETG modes whereas figure 21(\textit{c}) also shows instability to the KBM in the range $k_y\rho_{\rm{i}} \in [0.01, 0.7]$. These equilibria are much more unstable than the low-$\beta$ ones in the inner core but they are much more stable as we move towards the outer core with negative triangularity, high-$\beta$\footnote{We have not discussed the problem of accessibility of the high-$\beta$ state. Locally accessing these equilibria at constant magnetic shear while maintaining KBM stability is discussed in a recent work by~\citep{davies2022kinetic}.} equilibria showing the best characteristics. We believe that the outer-core stability is due to a large Shafranov shift and strong shaping. We also argue that negative triangularity has better characteristics than positive triangularity due to stronger shaping, i.e., ``squareness'' that we discussed in figure~\ref{fig:VMEC-Miller-fit}. It is also interesting to note that for the negative triangularity equilibria, the growth rate trend matches that of the ideal-ballooning stability in figure~\ref{fig:negtri_ball} --- less stable towards the core, more stable towards the edge. Since outer-core or edge transport is usually a limiting factor in experimental low-$\beta$ equilibria, these equilibria may be a novel alternative to realize higher-power devices.

\section{Summary \& Conclusions}
\label{sec:conclusions}
We began this work by exploring an analytical asymptotic procedure for solving the Grad-Shafranov equation and obtaining global axisymmetric equilibria in~\S\ref{sec:equilibrium}. We showed, in~\S\ref{sub:HAC_lim}, that the analytical method had several limitations, mostly related to the smoothness of the solutions, that made a local stability analysis infeasible. To remedy that, in~\S\ref{subsec:VMEC}, we generated numerical high-$\beta$ equilibria using the 3D equilibrium code \texttt{VMEC}. We produced different equilibria based for three beta values: low, intermediate and high ($\beta \sim 1$), each differing from the rest by at least an order or magnitude. We also chose two different boundary shapes with negative and positive triangularity. For each global equilibrium, we picked two local equilibria: one in the inner core and the other in the outer core. This gave us a total of $12$ local equilibria.

Upon ensuring that these equilibria did not suffer from the same limitations as the ones obtained using the asymptotic procedure, we first tested their stability to the infinite-$n$ ideal-ballooning mode in~\S\ref{sec:Ideal_ballooning_stability}. Since the ideal-ballooning mode is an MHD instability, we chose the local equilibria such that all the high and low-$\beta$ equilibria are ideal-ballooning stable. We also elucidated the idea of varying the magnetic shear and pressure gradient self-consistently, a.k.a, the Greene-Chance analysis in~\S\ref{subsec:Greene-Chance}. This powerful technique enabled us to know how far the local equilibria are from the region of marginal stability.

In~\S\ref{sec:microstability}, we briefly explained the linear, collisionless, gyrokinetic model. In~\S\ref{sec:ITG-study} and~\S\ref{sec:TEM-study}, we use this to study the stability of all the local equilibria to the two most virulent electrostatic modes of turbulence: ITG and TEM. We found a clear inverse relationship between the beta value and the growth rates --- increasing the beta value stabilized both the ITG mode and the TEM. Using a Greene-Chance analysis, we also scanned the maximum ITG, TEM growth rates vs the temperature and density gradient scale lengths, respectively. This was important to make sure that these equilibria are not ``stiff'', i.e., the growth rates do not increase sharply as the gradients exceed some threshold.
In~\S\ref{subsec:local-shear}, we explained how a large negative local shear resulting from a large Shafranov shift stabilizes the ITG mode and in~\S\ref{subsec:precession-drift} we showed how the reversal of the precession of the trapped electrons stabilizes the TEM throughout the whole range of gradient scale lengths.  

The effect of electromagnetic fluctuations can be important for all the equilibria, especially the intermediate and high-$\beta$ ones. To that end, in~\S\ref{sec:electromagnetic-GK} we performed an electromagnetic study of all the nominal equilibria.  We found that the stability trend seen for the electrostatic case did not hold after turning on electromagnetic effects. However, even though the high-$\beta$ equilibria were more unstable than the low-$\beta$ ones due to collisionless-MTMs and EM-ETGs in the inner core, they were much more stable than low-$\beta$ in the outer core. 
We found that negative triangularity, high-$\beta$ equilibria were stable to the KBM. We believe that this is due to the strong shaping (``squareness'') of the negative triangularity high-$\beta$ equilibria. This indicates that turbulent transport may flatten the pressure gradient in the core but may not significantly affect the pressure gradient towards the edge for the high-$\beta$ equilibria.

This work suggests many promising avenues for future research. One could repeat the same study with a wider range of input parameters to ascertain that the microstability trends we have found are not strongly dependent on the input parameters. An important path to explore would be to map the trajectory of these high-$\beta$ equilibria, starting from a low-beta equilibrium in the $\hat{s}-\alpha_{\mathrm{MHD}}$ space such that it is always stable to major sources of disruption. This would guarantee a high-$\beta$ operation free from any ideal-MHD instablilities. Once such a path is found, the final step would be to do a microstability calculation followed by a transport calculation to obtain the evolution of the density and temperature profiles and repeat the stability study for the steady-state temperature and density profiles.

\textbf{Acknowledgments}
This work was supported by the US Department of Energy, Office of Science, Office of Fusion Energy Sciences under Award Numbers DE-SC0018429 and DE-FG02-93ER54197. This research used resources of the National Energy Research Scientific Computing Center, a DOE Office of Science User Facility.
D.D. was also supported by TDoTP programme grant (EP/R034737/1). One of the authors, R.G., would like to thank Dr Matt Landreman for help setting up the \texttt{VMEC} code and providing the necessary input and post-processing files, Dr Mike Martin for sharing his \texttt{VMEC} to \texttt{GX} interface that helped R.G. in creating an improved interface between \texttt{VMEC} and \texttt{GS2}. The authors would like to thank Prof. Pierre Gourdain for providing us with the first high-$\beta$ equilibrium and Dylan Langone for doing the initial work that led to these investigation.

\appendix
\section{Straight-field-line $\theta$ calculation}
\label{app:st_field_line}
We briefly describe the process of calculating the straight-field-line $\theta$. We start with the angle $\theta_{\mathrm{geo}}$ which is be the geometric angle of a point on a flux surface measured about the magnetic axis. We assume the toroidal angle $\phi$ to be unchanged. We also assume that we know the various Jacobians $\mathcal{J}_{\mathrm{geo}} = (\boldsymbol{B}\cdot \bnabla \theta_{\mathrm{geo}})^{-1}$ and $\mathcal{J} = (\boldsymbol{B}\cdot \bnabla \theta)^{-1}$. The transformation relation will then be
\begin{equation}
    \theta = \theta_{\mathrm{geo}} - \tilde{\nu}(\psi,  \theta_{\mathrm{geo}}),
\end{equation}
where $\tilde{\nu}$ is a periodic function of $\theta_{geo}$ and  $\theta$ and $\tilde{\nu}(\psi, \pm \pi) = 0$ for consistency. Applying the $\boldsymbol{B}\cdot\bnabla$ operator
\begin{equation}
\begin{gathered}
    \boldsymbol{B}\cdot\bnabla \theta =  \boldsymbol{B}\cdot\bnabla \theta_{\mathrm{geo}} -  \boldsymbol{B}\cdot \bnabla\tilde{\nu}.
\end{gathered}
\label{eqn:straight-field-theta-appendix}
\end{equation}
A special property of the straight-field-line theta is that $q(\psi) = \boldsymbol{B}\cdot\bnabla\phi/\boldsymbol{B}\cdot\bnabla\theta$. This implies $\boldsymbol{B}\cdot \bnabla\theta = F/q R ^2$. Using this and dividing equation $\eqref{eqn:straight-field-theta-appendix}$ by $\boldsymbol{B}\cdot \bnabla\theta$ and integrating in $\theta_{\mathrm{geo}}$ from $-\pi$ to $\theta_{\mathrm{geo}}$
\begin{equation}
\begin{gathered}
    \tilde{\nu}(\theta_{\mathrm{geo}}) =  \theta_{\mathrm{geo}} - \int_{-\pi}^{\theta_{\mathrm{geo}}}\frac{q R^2}{F \mathcal{J}_{\mathrm{geo}}} d\theta, \\
    \theta  =  \int_{-\pi}^{\theta_{\mathrm{geo}}}\frac{q R^2}{F \mathcal{J}_{\mathrm{geo}}} d\theta. 
\end{gathered}
\end{equation}
Since all the equilibria we are studying are up-down symmetric, we can also assert $\tilde{\nu}(\psi, -\theta_{\mathrm{geo}}) = -\tilde{\nu}(\psi, \theta_{\mathrm{geo}})$ which means $\tilde{\nu}(\psi, 0) = 0$ and 
\begin{equation}
    \theta  =  \int_{0}^{\theta_{\mathrm{geo}}}\frac{q R^2}{F \mathcal{J}_{\mathrm{geo}}}\,  d\theta. 
\end{equation}

\section{Analytical solution of $\beta \sim 1$ equilibria}
\label{app:Hsu_et_al_profiles}
\label{app:Hsu-et-al-procedure}
To construct the $\beta \sim 1$ equilibrium in Hsu et. al's work, one has to start with analytical pressure, boundary shape profiles and the value of the poloidal field at $Z=0$:
\begin{equation}
    \begin{gathered}
    p(\hat{R}(\psi)) = p_1\left(1 - \frac{p_2(R_{\mathrm{max}}-\hat{R})^2 + p_3(R_{\mathrm{max}} - \hat{R})^3 + p_4(R_{\mathrm{max}}-\hat{R})^4}{p_2(R_{\mathrm{max}}-R_{\mathrm{min}})^2 + p_3(R_{\mathrm{max}} - R_{\mathrm{min}})^3 + p_4(R_{\mathrm{max}}-R_{\mathrm{min}})^4}\right),\\
    l_{\delta < 0}(\hat{R}) = (R-R_{\mathrm{min}})^{0.5}(R_{\mathrm{max}}-R)^{0.5}\left(\frac{a_l}{[R_{\mathrm{max}} - R + b_l(R_{\mathrm{max}}-R_{\mathrm{min}})]^{c_l}}\right)^{0.5},\\
    \frac{1}{\hat{R}}\frac{d \psi}{d \hat{R}} = \frac{l(\hat{R})}{\hat{R}}[a_{\psi} + b_{\psi}(\hat{R}-R_{\mathrm{min}}) + c_{\psi}(\hat{R}-R_{\mathrm{min}})^2],
    \end{gathered}
\end{equation}
where $l_{\delta<0}$ is a function used to generate the boundary shapes corresponding to positive triangularity equilibria. After choosing the value of the coefficients, we pick a given $\hat{R}(\psi)$ and using~$\eqref{eqn:xi-final-formula}$ calculate the perpendicular distance $\xi$ as a function of $\hat{R}$ and $R$. Next, we use~$\eqref{eqn:Core_and_BL_solution}$ to construct the core and boundary layer segments and join them at their intersection point. For figure~$\ref{fig:Hsu-VMEC-comparison}$ in this work which is figure $6$ in Hsu et. al, the values of the coefficients are given below
\begin{equation}
    \begin{gathered}
    (p_1, p_2, p_3, p_4) = (0.5, 1.5, 0, 0),\\
    (a_l, b_l, c_l) = (0.8, 0.05, 0.85),\\
    (a_{\psi}, b_{\psi}, c_{\psi}) = (0.152, 0.022, 0).
    \end{gathered}
\end{equation}

\section{Greene-Chance's method}
\label{app:Greene-Chance-method}
This appendix explains the Greene-Chance method, a powerful technique that allows one to vary the pressure and current gradients for a local equilibrium. We will start by defining the local coordinate system first developed by~\citet{mercier-luc-formalism}. After that, we will expand the Grad-Shafranov equation in those coordinates and use it to obtain other important relations, namely the gradients of $q, B$, and $\alpha$ around a surface. Finally, we explain how the derived relations can be used to vary a local equilibrium. All the lengths in the following calculations are normalized using $a_{\rm{N}}$ and the magnetic fields using $B_N$. For appendices C.1 and C.2, we define $\rho$ to be the normalized radial distance from a flux surface.
\subsection{Mercier-Luc local coordinate system}
\begin{figure}
    \centering
    \includegraphics[width=0.5\linewidth]{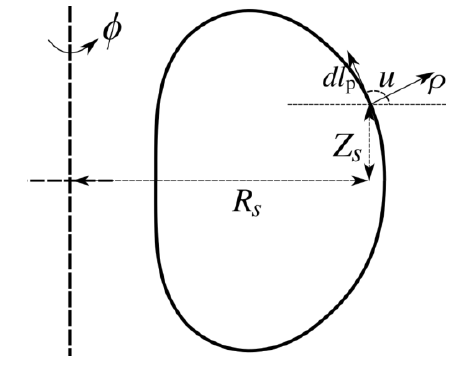}
    \caption{shows the local coordinate system of Mercier and Luc. The poloidal arc length $l_{\rm{p}}$ is measured in a counter-clockwise sense from the outboard side. We define $\theta = \theta_{\rm{geo}} = 0$ at the farthest point from the symmetry axis.}
    \label{fig:ML-schematic}
\end{figure}

The Mercier-Luc(ML) coordinate system is $(\rho, \phi, l_{\rm{p}})$ where $\rho$ is the normalized perpendicular distance from a point on the flux surface, $\phi$ is the cylindrical azimuthal angle, and $l_{\rm{p}}$ is the normalized poloidal arc-length. In these local coordinates we can write, the cylindrical $(R, \phi, Z)$ coordinates as
\begin{equation}
\begin{gathered}
    R = R_0 + \rho \sin(u(l_p)) + \int_{0}^{l_{\rm{p}}} \cos(u)\, d{l^{'}_{\rm{p}}},\\
    Z = Z_0 + \rho \cos(u(l_p)) + \int_{0}^{l_{\rm{p}}} \sin(u)\, d{l^{'}_{\rm{p}}},\\
\end{gathered}
\label{eqn:Cylindrical-to-ML-transformation}
\end{equation}
where $R_0, Z_0$ are the normalized $R, Z$ values on the outboard mid-plane of the flux surface of interest and the angle $u$ is defined as shown in figure~\ref{fig:ML-schematic}. We also define
\begin{equation}
    \begin{gathered}
        R_s \equiv R(\rho = 0, l_{\rm{p}}) = R_0 + \int_{0}^{l_{\rm{p}}} \cos(u) d{l^{'}_{\rm{p}}},\\
        Z_s \equiv Z(\rho = 0, l_{\rm{p}}) = Z_0 + \int_{0}^{l_{\rm{p}}} \sin(u) d{l^{'}_{\rm{p}}},
    \end{gathered}
\end{equation}
as the on-surface cylindrical coordinates. The azimuthal angle $\phi$ is the same for both cylindrical and ML coordinates. Using equation $\eqref{eqn:Cylindrical-to-ML-transformation}$ and after choosing a sign for the curvature, we can define the curvature
\begin{equation}
    \frac{1}{R_{\rm{c}}} = \frac{du}{dl_{\rm{p}}}.
\end{equation}
where $R_{\rm{c}}$ is the radius of curvature.
Using the coordinate transformation relation $\eqref{eqn:Cylindrical-to-ML-transformation}$, we can write
\begin{equation}
\setlength{\arraycolsep}{0pt}
\renewcommand{\arraystretch}{1.3}
\left[ 
\begin{matrix}
	dR\\[2mm]
	d\phi\\[2mm]
	dZ
\end{matrix}	
\right] = 
 \left[ 
	 \begin{array}{ccc}
		  \sin(u) \quad & 0 &\quad  \left(1+\frac{\rho}{R_{\rm{c}}}\right) \cos(u)\\[2mm]
		       0  \quad & 1 &\quad 0 \\[1mm]
		  -\cos(u)\quad & 0 &\quad \left(1+\frac{\rho}{R_{\rm{c}}}\right) \sin(u) 	
	  \end{array}
 \right]
 \left[
	 \begin{matrix}
		 d\rho\\[2mm]
		 d\phi\\[2mm]
		 dl_{\rm{p}}
	 \end{matrix}
 \right]\\ 
\end{equation}
This matrix can be inverted to obtain the transformation from ML to cylindrical coordinates. Using this transformation, we can write the transformation Jacobian 
\begin{equation}
\tilde{\mathcal{J}} = ((\bnabla\rho \times \bnabla\phi)\cdot \bnabla l_{\rm{p}})^{-1} = R \left(1 + \frac{\rho}{R_{\rm{c}}}\right).
\end{equation}
\subsection{Expanding the Grad-Shafranov equation locally}
The Grad-Shafranov equation 
\begin{equation}
    R^2 \bnabla \cdot\left(\frac{\bnabla \psi}{R^2}\right) = -R^2 \frac{dp}{d\psi} - F\frac{dF}{d\psi}, 
\end{equation}
can be written in the ML coordinate system as
\begin{equation}
    \frac{R}{(1+\rho/R_c)}\left[ \frac{\partial }{\partial l_{\rm{p}}}\left(\frac{(1+\rho/R_{\rm{c}})}{R} \frac{\partial \psi}{\partial l_{\rm{p}}}\right) + \frac{\partial}{\partial \rho}\left(\frac{(1+\rho/R_{\rm{c}})}{R} \frac{\partial \psi}{\partial \rho} \right)\right] = -R^2 \frac{dp}{d\psi} - F\frac{dF}{d\psi}. 
    \label{eqn:Local-Grad-Shafranov-equation}
\end{equation}
To obtain the local dependence of $\psi$ to $dF/d\psi$ and $dp/d\psi$, we write $\psi$ as an asymptotic series in terms of the normalized radial distance $\rho$
\begin{equation}
    \psi = \psi_0 + \rho \psi_1(l_{\rm{p}}) + \rho^2 \psi_2(l_{\rm{p}}) + \textit{O}(\rho^3),
\end{equation}
and define
\begin{equation}
    \psi_1 = \lim_{\rho \rightarrow 0} \left(\frac{\psi-\psi_0}{\rho}\right) = -R_{s} B_{{\rm{p}}s},\quad  \psi_1 < 0,\, B_{{\rm{p}}s} > 0. 
\end{equation}
where $B_{{\rm{p}}s}$ is the on-surface poloidal magnetic field. Another way to write the above relation is to say ${\bnabla}\psi\lvert_{\rho = 0} = -\psi_1 \bnabla\rho$. Using this asymptotic expansion, we can also write the following Taylor series expansions
\begin{equation}
\begin{gathered}
    F(\psi) = F(\psi_0) + \frac{\psi-\psi_0}{1!}\frac{d F}{d \psi}\bigg\lvert_{\psi_0} + \ldots = F(\psi_0) - \rho R_{\rm{s}} B_{{\rm{p}}s} \frac{d F}{d \psi}\bigg\lvert_{\psi_0} + \ldots ,\\
    F^{'}(\psi) = F^{'}(\psi_0) + \frac{\psi-\psi_0}{1!}\frac{d F^{'}}{d \psi}\bigg\lvert_{\psi_0} + \ldots  = F^{'}(\psi_0) - \rho R_{\rm{s}} B_{{\rm{p}}s}\frac{d F^{'}}{d \psi}\bigg\lvert_{\psi_0} + \ldots ,\\
    p(\psi) = p(\psi_0) + \frac{\psi-\psi_0}{1!}\frac{d p}{d \psi}\bigg\lvert_{\psi_0}  + \ldots = p(\psi_0) - \rho R_{\rm{s}} B_{{\rm{p}}s}\frac{d p}{d \psi}\bigg\lvert_{\psi_0} + \ldots ,\\
    q(\psi) = q(\psi_0) + \frac{\psi-\psi_0}{1!}\frac{d q}{d \psi}\bigg\lvert_{\psi_0} + \ldots = q(\psi_0) - \rho R_{\rm{s}} B_{{\rm{p}}s}\frac{d q}{d \psi}\bigg\lvert_{\psi_0} + \ldots ,
    \end{gathered}
\end{equation}
where the prime denotes a derivative with respect to $\psi$. After substituting the Taylor series expansions into the local Grad-Shafranov equation \eqref{eqn:Local-Grad-Shafranov-equation}, we get
\begin{equation}
    \psi_2 = -\frac{1}{2}\left[R_{s} B_{{\rm{p}}s}\left(\frac{\sin(u)}{R_s} - \frac{1}{R_{\rm{c}}}\right) + R_{s}^2\, p^{'}(\psi_0)  + F(\psi_0) F^{'}(\psi_0)\right].
\end{equation}
Next, we expand the relation $\eqref{eqn:Safety-factor-definition}$ about a flux surface to get
\begin{equation}
\begin{split}
 \frac{dq}{d\psi} &= F^{'}\left(\frac{q}{F} + \frac{q F}{2\pi} \oint \frac{1}{(R_s B_{{\rm{p}}s})^2} d\theta\right) + \frac{p^{'}q}{2\pi} \oint \frac{d\theta}{ B_{{\rm{p}}s}^2}\\
  &+ \frac{q}{2\pi}\oint \frac{2\, d\theta}{ R_s B_{{\rm{p}}s}} \left(\frac{\sin(u)}{R_s} - \frac{1}{R_c} \right).
 \end{split}
\end{equation}
This gives us an algebraic equation defining $q^{'}$ in terms of $F^{'}$ and $p^{'}$
\begin{equation}
\begin{split}
    \frac{1}{q}\frac{dq}{d\psi} &= F^{'}\left(\frac{1}{F} +  \frac{F}{2\pi} \oint d\theta \frac{1}{(R_s B_{{\rm{p}}s})^2} \right) + \frac{p^{'}}{2\pi} \oint \frac{d\theta}{ B_{{\rm{p}}s}^2}\\
    &+ \frac{1}{2\pi}\oint \frac{2\, d\theta}{ R_s B_{\rm{p}}s} \left(\frac{\sin(u)}{R_s} - \frac{1}{R_c} \right),
\end{split}
\label{eqn:full-magic-formula}
\end{equation}
written more compactly as
\begin{equation}
    \frac{1}{q}\frac{dq}{d\psi} = F^{'} a_{s,\mathrm{full}} + p^{'} b_{s,\mathrm{full}}+  c_{s,\mathrm{full}},
    \label{eqn:Bishop-shear-magic-formula}
\end{equation}
where $a_{s,\mathrm{full}}, b_{s,\mathrm{full}},$ and $c_{s,\mathrm{full}}$ are three constants obtained from doing the respective integrals in~\eqref{eqn:full-magic-formula}. Using the ML coordinates, we can also expand the magnetic field strength around a flux surface
\begin{equation}
\begin{split}
    B^2 &= \frac{F^2}{R^2} + \left(\frac{1}{R}\frac{d\psi}{d\rho}\right)^2, \\
     &= B_s^2\left[ 1 + \frac{2\rho}{B_s^2}\left(-\frac{B_{{\rm{p}}s}^2}{R_{\rm{c}}} + R_s B_{{\rm{p}}s}p^{'} - \frac{F^2}{R_s^3}\sin(u)\right)\right],
\end{split}
\end{equation}
which gives us the local, radial gradient of the magnetic field
\begin{equation}
    \frac{\partial B}{\partial \rho} = \frac{1}{B_s}\left(-\frac{B_{{\rm{p}}s}^2}{R_{\rm{c}}} + R_s B_{{\rm{p}}s}p^{'} - \frac{F^2}{R_s^3}\sin(u)\right).
    \label{eqn:dB-drho}
\end{equation}
To obtain all the geometric coefficietns, we also need various gradients of the field line label $\alpha$. To that end, we can write
\begin{equation}
    \alpha = \phi + S(\rho, l_{\rm{p}}) = \phi - q\, \theta,
\end{equation}
and find $S = S(\rho, l_{\rm{p}})$ by solving the equation
\begin{equation}
\boldmath{B}\cdot\bnabla \alpha = 0, 
\label{eqn:B-dot-grad-S}
\end{equation}
in ML coordinates. To solve equation~\eqref{eqn:B-dot-grad-S}, we write $S(\rho, l_{\rm{p}})$ as an asymptotic series in $\rho$ 
\begin{equation}
    S = S_0(l_{\rm{p}}) + \rho S_1(l_{\rm{p}}) + \textit{O}(\rho^2),  
\end{equation}
and equation $\eqref{eqn:B-dot-grad-S}$ becomes
\begin{equation}
    \frac{F}{R^2} + \frac{1}{R(1 + \rho/R_{\rm{c}})} \frac{\partial \psi}{\partial \rho}\frac{\partial \alpha}{\partial l_{\rm{p}}} - \frac{1}{R(1 + \rho/R_{\rm{c}})} \frac{\partial \psi}{\partial l_{\rm{p}}}\frac{\partial \alpha}{\partial \rho} = 0.
\end{equation}
The lowest order solution yields
\begin{equation}
    S_0 = \int dl_{\rm{p}} \frac{F(\psi_0)}{R_s^2 B_{{\rm{p}}s}} + f(\rho),
\end{equation}
For axisymmetric equilibria, all the field lines are identical which means we can choose $f(\rho) = 0$ without loss of generality. The next order solution gives us
\begin{equation}
\begin{split}
    S_1 &= -R_s B_{{\rm{p}}s} \left[ F^{'}\left(\frac{q \theta }{F} +  F \int_{0}^{\theta} d\theta \frac{q}{(R_s B_{{\rm{p}}s})^2} \right) \right.  \\
        & \left. +\, p^{'} \int_{0}^{\theta}  d\theta\frac{q}{ B_{{\rm{p}}s}^2} + \int_{0}^{\theta} d\theta \frac{2 q}{ R_s B_{{\rm{p}}s}} \left(\frac{\sin(u)}{R_s} - \frac{1}{R_{\rm{c}}} \right)\right],
\end{split}
\end{equation}
\begin{equation}
    \frac{\partial \alpha}{\partial \rho} = S_1 = -R_s B_{{\rm{p}}s}\left(F^{'} a_s + p^{'} b_s + c_s\right).
    \label{eqn:Bishop-alpha-magic-formula}
\end{equation}
This completely defines the quantity $\bnabla \alpha$. Using~\eqref{eqn:Bishop-shear-magic-formula},~\eqref{eqn:dB-drho},~\eqref{eqn:Bishop-alpha-magic-formula} we can calculate all the geometric coefficients\footnote{To include the effect of $\theta_0$, we just have to change the lower limit of integration in~\eqref{eqn:Bishop-alpha-magic-formula}.} needed for a local stability analysis. As an example, we obtain an analytical formula for the normalized local magnetic shear along the field line in a flux surface
\begin{equation}
    \nu = -\boldsymbol{B}\cdot{\bnabla}\left(\frac{{\bnabla}\alpha\cdot {\bnabla}\psi}{|{\bnabla}\psi|^2}\right).
\end{equation}
Using the following equation 
\begin{equation}
\boldsymbol{B}\cdot{\bnabla}\left(\frac{{\bnabla}\alpha \cdot {\bnabla}\psi}{|{\bnabla}\psi|^2} \right)=  (\boldsymbol{B}\cdot {\bnabla}\theta) \frac{\partial}{\partial \theta}\left(\frac{d\rho}{d\psi}\frac{\partial \alpha}{\partial \rho}\right), \\
\end{equation}
we obtain the local shear 
\begin{equation}
    \nu = -\left[\left(\frac{q F' }{F} +  F^{'} F \frac{q}{(R_s B_{{\rm{p}}s})^2}\right) + \frac{qp^{'}}{ B_{{\rm{p}}s}^2} + \frac{2 q}{ R_s B_{{\rm{p}}s}} \left(\frac{\sin(u)}{R_s} - \frac{1}{R_{\rm{c}}} \right)\right].
\end{equation}
One can verify that $dq/d\rho = (\partial \psi/\partial \rho) \int_{0}^{2\pi} d\theta/($$\boldmath{B}\bcdot \bnabla \theta$$)\, \nu$.

\subsection{Variation of a local axisymmetric equilibrium}
After redefining $\rho = \sqrt{\chi/\chi_{\rm{LCFS}}}$, equation $\eqref{eqn:Bishop-shear-magic-formula}$ can be written as 
\begin{equation}
\begin{gathered}
    \frac{\hat{s}}{\rho} \equiv \frac{1}{q}\frac{dq}{d\rho} = \frac{dF}{d\rho}\, a_{\rm{s,full}} + \frac{dp}{d\rho}\, b_{\rm{s,full}} +c_{\rm{s,full}},\\
\end{gathered}
\label{eqn:shat-Bishop-formula}
\end{equation}
where  $a_{\rm{s,full}}, b_{\rm{s,full}}, c_{\rm{s,full}}$ are constants. This equation implies that on a given surface, we can vary the pressure gradient $dp/d\rho$ and $\hat{s}$ of a local equilibrium independently by a finite amount as long as we adjust $dF/d\rho$ such that the equation $\eqref{eqn:shat-Bishop-formula}$ is satisfied. Once $dp/d\rho, dq/d\rho$, and $dF/d\rho$ are fixed, we have fully defined a local equilibrium. Greene and Chance first used this idea to perform an $\hat{s}-\alpha_{\mathrm{MHD}}$ analysis. We are going to use this for the ballooning and gyrokinetic stability analyses.

\section{Newcomb's criterion}
\label{app:Newcomb's-criterion}
\citet{newcombcriterion}, in his analysis of a diffuse screw pinch, described a method to infer the stability of a system to incompressible ideal-MHD modes. He reduced the ideal-MHD energy principle to 
\begin{equation}
    W = \frac{\pi}{2}\int_{r_1}^{r_2} dr \left[f\left(\frac{d\xi}{dr}\right)^2 + g\xi^2\right],
\end{equation}
where $r$ is the distance from the center, $\xi(r)$ is the radial perturbation and $f(r), g(r)$ are functions dependent on the equilibrium. One can write the kinetic energy associated with the perturbation as 
\begin{equation}
    T = \frac{\pi\, \omega^2}{2}\int_{r_1}^{r_2} dr\,  \xi^2.
\end{equation}
Combining the potential and the kinetic energy, we can write the Lagrangian $L = T - U$ and get the Euler-Lagrange equation corresponding to $\delta L = 0$
\begin{equation}
    (f \xi^{'})^{'} - g\xi = \omega^2 \xi.
\label{eqn:pinch-equation}
\end{equation}
This equation a self-adjoint, second-order, eigenvalue ordinary differential equation(ODE). In his paper, Newcomb explored the marginally stable ODE,
\begin{equation}
    (f \xi^{'})^{'} - g\xi = 0.
\label{eqn:marginal-pinch-equation}
\end{equation}
For the marginally stable equation of this form, Newcomb's theorem and the associated corollary is given below:
\begin{theorem}
If $r_{1}$ and $r_{2}$ are nonsingular points of the same independent sub-interval, and if the nontrivial Euler-Lagrange solutions that vanish at $r_1$ also
vanish at some point $r_0$ between $r_1$ and $r_2$ , then for any Euler-Lagrange solution
$\xi_0(r)$ there exist functions $\xi(r)$ with the same boundary values and with $W(r_1 ,
r_2; \xi) < W(r_1, r_2; \xi_0)$
\end{theorem}

\begin{corollary}
There exists a $\xi(r)$ that makes $W(r_1, r_2; \xi)$ negative and satisfies the boundary conditions $\xi(r_1) = \xi(r_2) = 0$
\end{corollary}
This means that upon integrating equation $\eqref{eqn:marginal-pinch-equation}$ with a test function $\xi$, such that $\xi(r_1) = 0$, if $\xi$ crosses the zero line at any other point, then there must exist an eigenfunction $\tilde{\xi}$ satisfying equation $\eqref{eqn:pinch-equation}$ such that $W < 0$, implying that $\omega^2 < 0$. In other words, the system will have a growing eigenvalue and will become unstable. 

Even though Newcomb derived this theorem for an ODE that is integrated in the radial direction, the principle can be extended to any second-order, self-adjoint, eigenvalue ODE. Hence, we can use it here for the ideal-ballooning equation \eqref{eqn:ballooning-equation}.

\bibliographystyle{jpp}

\bibliography{main}

\end{document}